\documentclass[preprint,amsmath,aps,amssymb,pra,showpacs]{revtex4-1}
\usepackage{graphicx}
\DeclareMathOperator{\csch}{csch}
\begin{document}

\title{Fourth-Order Master Equation for a Charged Harmonic Oscillator Interacting with the Electromagnetic Field}

\author{Arzu Kurt}
\author{Resul Eryigit}
\email{resul@ibu.edu.tr}
\affiliation{Department of Physics, Abant Izzet Baysal University, Bolu, Turkey}

\date{\today}

\begin{abstract}
The master equation for a charged harmonic oscillator coupled to an electromagnetic reservoir is investigated up to fourth-order in the interaction strength by using Krylov averaging method. The interaction is in the velocity-coupling form and includes a diamagnetic term. Exact analytical expressions for the second, the third, and the fourth-order contributions to mass renormalization, decay constant, normal and anomalous diffusion coefficients are obtained for the blackbody type environment. It is found that, generally, the third and the fourth order contributions have opposite sign when their magnitudes are comparable to that of the second order one.
\end{abstract}
\maketitle
\section{Introduction}

Research activity in dissipation and decoherence of quantum systems has 
accelerated in the last two decades with developments in quantum information 
technologies, progress in detection and control of ultra-cold atom and 
ions, atom lasers and photonic band-gap 
materials as well as the research on the role of decoherence in quantum-to-classical 
transition~\cite{lewenstein2012,massignan2015,breuer2002theory,schlosshauer2007decoherence,weiss1999quantum}.
Theoretical efforts to investigate dissipation and decoherence phenomena due to interaction between a quantum system and its environment use either exact or perturbative treatments. While exact approaches can be used to treat a 
restricted class of problems for the whole interaction parameter range, the 
perturbative treatments can be applied to almost all problems, but only for a 
restricted range of parameters which is, generally, the weak coupling regime. 
A plethora of different perturbation based formalisms have been developed to treat such system-bath problems over the last six decades~\cite{breuer2002theory,schlosshauer2007decoherence,weiss1999quantum}. 
The most prominent among them are Nakajima-Zwanzig projector 
operator method, path-integral based influence functional method~\cite{hu1992}, self-consistent hybrid schemes, Monte Carlo methods and hierarchical approaches~\cite{tanimura2006,liu2014}. Most of the perturbative approaches express the equation of motion of the reduced density matrix of the system as a quantum master equation (QME) which contains second-order interaction terms. 

One way to increase the validity range of the perturbative methods might be increasing the order of 
master equation. Actually, it is shown by Fleming  and Cummings~\cite{fleming2011} that an 
order-$2n$ accuracy in the full-time solutions of master equations requires an order-$(2n+2)$ master 
equation. A number of groups investigated the effects of the fourth and higher order terms in detail
~\cite{Skinner87,laird1991,PhysRevE.55.2328,1408,ford1996,
reichman1996, Breuer1999,Breuer19992, breuer200136,jang2002,doll2008243,tanimura2008,fleming2011,singh2012,liu2014}. 
In particular, Reichman et. al. have derived analytical expressions for dynamics of a harmonic oscillator in contact with various types of baths in the low temperature limit~\cite{reichman1996}. Also, a physically sound fourth-order correction to Redfield equation has been worked out by Laird, Budimir and Skinner~\cite{laird1991}. Jang, Cao and Silbey~\cite{jang2002} have derived a 
fourth-order QME in both time local and nonlocal forms for a general system 
Hamiltonian and used it to study motion of a particle in a continuous potential 
field and the dynamics of a two-level system coupled to a bath. Fourth-order 
corrections were found to cause a potential renormalization for the first 
problem while they introduced additional coherence for the later.
In a series of papers, 
Breuer et. al.~\cite{Breuer1999,Breuer19992,breuer200136} have examined, in detail, the effects of 
including the fourth and higher order terms in time convolutionless master equation for the damped Jaynes-Cummings model (JCM), atom laser and the damped harmonic oscillator. It was found that at weak and moderate coupling, the fourth order QME was in good agreement with the exact solution for the JCM while for the intermediate coupling regime of atom lasers, one needs to 
include non-Markovian as well as the fourth- and sixth-order terms to the master equation to adequately describe the dynamics of the 
system. Singh and Brumer investigated the 
validity of the of the second-order Markovian Redfield theory for 
reorganization energy and decay rates of photon autocorrelation functions in dimer systems~\cite{singh2012}.  Liu et. 
al.~\cite{liu2014} used hierarchical equation of motion method for the 
spin-boson problem and showed that fourth-order corrections are important for 
the intermediate coupling regime. Mavros and Voorhis considered fourth-order 
corrections to the memory kernel of the generalized master equation of the 
spin-boson problem and found that a numerically-exact solution was possible 
when the system-bath coupling is weak~\cite{Mavros2014}. Interestingly, it was found that 
the non-perturbative treatment of certain system-bath problems 
is equivalent to the second-order perturbative master equations and 
including higher than second order terms would be detrimental 
for such systems~\cite{tanimura2008,doll2008243}.

Dissipative dynamics of a charged particle in an electromagnetic field, whether 
a coherent laser field or an incoherent blackbody radiation, have been examined 
by several 
groups~\cite{ford88,marathe1989,ford1996,li1996,dattagupta1997,bao2005,bai2005,
kalandarov2007,pachon2013,kumar2014} because of its relevance to damped transport in solid-state 
systems. Bao and Bai~\cite{bao2005,bai2005} have considered the the motion of a single electron atom interacting with an electromagnetic field in the dipole coupling 
approximation and obtained a ballistic diffusion for the particle in the  position space. 
Kalandarov et. al.~\cite{kalandarov2007} have shown that dissipation of 
collective energy of a charged oscillator linearly coupled to a heat bath could be controlled by an 
external field. Pachon and  Brumer~\cite{pachon2013} investigated the effect of blackbody radiation 
on the coherence of the dynamics  of light-harvesting molecules in photosynthesis and showed that the blackbody radiation would not enhance the coherence of the system. 

Ford, Lewis, and O'Connell derived a master equation for a charged oscillator weakly coupled to an electromagnetic field by using Krylov averaging method which can be improved by including higher-order interaction terms in a systematical way~\cite{ford1996}. The model has several interesting features, such as use of velocity-coupling for the system-environment interaction which is not as widely studied as the position-coupling and keeping the  diamagnetic term of the interaction Hamiltonian which leads to the nonzero third order contributions to the master equation which are absent for most of similar studies. In the current work, our aim is two-fold; we aim for exact analytical expressions for all coefficients of the derived master equation which are obtained in the limit of infinite environmental cutoff frequency in Ref.~\cite{ford1996}. Our findings indicate that the exact treatment is crucial for the accurate high frequency behavior of the diffusion coefficients. 
Second purpose of the present study is to derive analytical expressions 
for the fourth-order corrections to the master equation of the charged 
oscillator in an electromagnetic field and delineate the convergence range of interaction parameters.  

The structure of this paper is as follows. We will introduce the model of the charged harmonic oscillator in contact with a blackbody radiation field in Section~\ref{model} where we provide a short review of the derivation of the master equation by Krylov averaging method. Analytical expressions for the second, the third and the fourth order corrections are, also, presented in Section~\ref{model} and discussed in Section~\ref{discus}. Section~\ref{conc} 
concludes the paper with a short review of the main findings. We provide the details of the derivation of the fourth order terms in the Appendix.

\section{\label{model}Model and Method}
 We consider a charged harmonic oscillator in a blackbody radiation field~\cite{ford1996,Bai05,ford1988}. The total Hamiltonian of the closed system formed by the oscillator and field can be written as:
\begin{equation}
\label{eq:Ham}
H=H_{S}+H_{R}+\epsilon H_{I},
\end{equation}
with
\begin{eqnarray}
H_{S}&=&\frac{p^{2}}{2\,M}+\frac{1}{2}K\,x^{2},\\ H_{R}&=&\sum\limits_{j}\left(\frac{p_{j}^2}{2\,m_{j}}+\frac{1}{2}m_{j}\,\omega_{j}^2\,q_{j}^2\right),\\ \epsilon H_{I}&=&\frac{p\,A}{M}+\frac{A^2}{2\,M},
\end{eqnarray} 
where $H_{S}$, $H_{R}$, and $\epsilon\,H_{I}$ describe the charged harmonic oscillator as the system, blackbody radiation field as the reservoir and the interaction between the velocity of the oscillator and the vector potential $A$ of the field, respectively. $K$ and $M$ are the force constant and the mass of the charged oscillator, while $m_{j}$, $p_{j}$, $q_{j}$ and $\omega_{j}$ describe  the environmental oscillators that stand for the blackbody field. $p$ and $x$ for the oscillator and $p_{j}$, $q_{j}$ of the reservoir obey the usual commutation rules as $\left[x,p\right]=i\hbar$, $\left[q_{i},p_{j}\right]=i\hbar\delta_{ij}$ and $\left[x,p_{j}\right]$=$\left[p,q_{j}\right]$=0. The vector potential of the field $A$ can be expressed as:
\begin{equation}
A=\sum_{j}m_{j}\,\omega_{j}\,q_{j}.
\end{equation}

The interaction between the oscillator and the field is assumed to be in velocity coupling type and contains both $p.A$ and a diamagnetic $A^{2}$ terms. It is well known that the Hamiltonian in Eq.~(\ref{eq:Ham}), can be transformed to a renormalized coordinate-coupling form by using Power-Zienau transformation~\cite{Babiker1974}. The diamagnetic interaction $A^{2}$ is generally neglected ~\cite{PhysRevA.35.4122,PhysRevA.43.57}, because its inclusion makes calculation of path-integral based master equation difficult. But its inclusion is shown to be important for the correct derivation of the partition function of a charged oscillator in blackbody field ~\cite{ford1988,PhysRevA.38.527} and we keep it here.

The influence of the environment on the system is determined by the spectral density $J(\omega)$ of the bath which is a measure of the number of bath oscillators at frequency $\omega$ and the coupling constant of the system-bath interaction. For the present case, we use the Drude corrected Ohmic spectral density:
\begin{equation}
\label{spectralDensity}
J\left(\omega\right)=m\tau_e\omega\frac{\Omega^2}{\omega^2+\Omega^2},
\end{equation}
\noindent where $\tau_e=2 e^2/(3 mc^3)=6.24\;10^{-24}$ s is the 2/3 of the time it takes light to traverse the classical radius of electron. $\Omega$ is the cutoff frequency of the bath modes which sets bath correlation time as $\tau_B=\Omega^{-1}$. There is an upper limit of the form $\Omega\le \tau_e^{-1}$ based on the causality arguments~\cite{OConnell2003}.

In the interaction picture, the unitary dynamics of the closed system formed by the oscillator and the radiation field is described by the von Neumann equation for the density matrix $\rho$ as:
\begin{equation}
 \label{eq:IntNeumann}
 \frac{\partial \rho\left(t\right)}{\partial t}=\frac{\epsilon}{i\, \hbar}\left[H_{I}\left(t\right),\rho\left(t\right)\right],
 \end{equation}
\noindent where the interaction picture transformation is taken with respect to $H_S$ and $H_R$ as
 \begin{subequations}
 \begin{eqnarray}
     \epsilon H_{I}\left(t\right)&=&e^{\frac{i\left(H_{S}+H_{R}\right)t}
      {\hbar}}\epsilon H_{I}e^{-\frac{i\left(H_{S}+H_{R}\right)t}{\hbar}}\label{eq:inthama}\\
       &=&\frac{p\left(t\right)A\left(t\right)}{m}+\frac{A^{2}\left(t\right)}{2\,m},\nonumber\\
A\left(t\right)&=&e^{\frac{i\,H_{R}\,t}{\hbar}}A\,e^{\frac{-i\,H_{R}\,t}{\hbar}}\label{eq:Arfunc}\nonumber\\
    &=&\sum\limits_j\{m_{j}\,\omega_{j}\,q_{j}\cos\left(\omega_{j}\,t\right)+p_{j}\sin\left(\omega_{j}\,t\right)\}
    ,\\
 \rho\left(t\right)&=&e^{\frac{i\left(H_{S}+H_{R}\right)t}{\hbar}}\rho\,e^{-\frac{i\left(H_{S}+H_{R}\right)t}{\hbar}}\label{eq:inthamb}.
\end{eqnarray}
\end{subequations}
 We assume that the bath and the oscillator are uncorrelated initially, that is 
 \begin{equation}
\label{eq:Iniden}
\rho\left(0\right)=\rho_{S}\left(0\right)\otimes\rho_{R},
\end{equation}
where $\rho_{S}=\mathrm{Tr}_R\{\rho\}$ is the reduced density matrix of the oscillator and $\rho_{R}$ is the density matrix of the reservoir which is assumed to be thermal at temperature $T$. $\mathrm{Tr}_R$ indicates partial trace over the bath degrees of freedom. 

There have been may attempts to solve Eq.~(\ref{eq:IntNeumann}) by using various different approximations, such as projector operator~\cite{ZPhysB,Yoon75}, path integral influence functional formulations~\cite{Feynman}. When the system-environment interaction is weak, Krylov averaging method can also be used to obtain a hierarchy of equations for the dynamics of the oscillator~\cite{ford1996}. For the self-containment of the present paper, we give a short overview of the method and the details of the derivation of the fourth order terms in the Appendix section.  

For ease of comparison with Ref.~\cite{ford1996}, 
we will rearrange the commutators on the right-hand side of Eq.~(\ref{eq:CofB3}) as 
\begin{eqnarray}
\label{eq:secOrd}
\frac{\partial \rho_{S}}{\partial t}=&&-\frac{1}{2\,m\,\hbar}\left(i\,A_{1}\left[p\left(t\right)^{2},\rho_{S}\right]+A_{2}\left[p\left(t\right),\left[p\left(t\right),\rho_{S}\right]\right]+m\,\omega_{0}\,A_{3}\left[p\left(t\right),\left[x\left(t\right),\rho_{S}\right]\right]\right.\nonumber\\
&&+\left.A_{4}\left(\left[p\left(t\right),\{p\left(t\right)-i\,m\,\omega_{0}\,x\left(t\right)\}\rho_{S}\right]+\left[\rho_{S}\{p\left(t\right)+i\,m\,\omega_{0}\,x\left(t\right)\},p\left(t\right)\right]\right)\right),
\end{eqnarray}
\noindent by defining 
\begin{eqnarray}
A_{1}&=&\frac{1}{\omega_{0}}\mathrm{Im}\left(T_{1}\right),\quad
A_2=\frac{1}{\omega_{0}}\left(\mathrm{Re}\left(T_{1}\right)
+\mathrm{Im}\left(T_{2}\right)\right),\nonumber\\
A_3&=&\frac{1}{\omega_{0}}\mathrm{Re}\left(T_{2}\right),\quad
A_4=\frac{1}{\omega_{0}}\mathrm{Im}\left(T_{2}\right)
\label{eq:defAis}.
\end{eqnarray}

At the second order, we need to evaluate $T^{(2)}_{1}$ and $T^{(2)}_{2}$ defined in equations~(\ref{T21}) and (\ref{T22}). Both of them can be calculated exactly by using the rational expansion of the hyperbolic cotangent
\begin{equation}
\coth\left(\beta\,\omega_{0}\right)=\frac{1}{\beta\,\omega_{0}}+\sum\limits_{n=1}^{\infty}\frac{2\,\beta\,\omega_{0}}{\beta^{2}\,\omega_{0}^{2}+n^{2}\,\pi^{2}}
\label{eq:coth}
\end{equation}
\noindent as follows:
\begin{subequations}
\begin{eqnarray}
T_{1}^{(2)}&=&f_{\omega_0}\left(\gamma\, \coth{\left(\beta\omega_0\right)}-i\, \omega_0\frac{\delta m}{m}\right)\label{eq:t21},\\
T_{2}^{(2)}&=&f_{\omega_0}\gamma\left(I-i \right)\label{eq:t22},
\end{eqnarray}
\end{subequations}
\noindent
\noindent where the mass renormalization factor $\delta m/m$ and the decay constant $\gamma$ are defined in Ref.~\cite{ford1996} as 
\begin{equation}
\label{eq:correction}
\frac{\delta m}{m}=\Omega\,\tau_e\quad \mathrm{and}\quad \gamma=\omega_{0}^{2}\,\tau_{e}\,,
\end{equation}
\noindent and 
\begin{equation}
I=-\frac{1}{\beta\,\Omega}+\frac{2}{\pi}\left(\mathrm{Re}
\left\{\psi_{0}\left[1-i\,\frac{\beta\,\omega_{0}}{\pi}\right]
\right\}-\psi_{0}\left[\frac{\beta\,\Omega}{\pi}\right]\right),
\label{eq:i1}
\end{equation}
\noindent where $\psi_{0}\left[x\right]$ is the digamma function and $\mathrm{Re}$ stands for the real part of its argument. One should note that $T_{1}^{(2)}$ and $T_{2}^{(2)}$ of equations~(\ref{T21}) and (\ref{T22}) correspond to $C$ and $S$ integrals of Ref.~\cite{ford1996}, respectively. They were evaluated at the $\Omega\rightarrow\infty$ limit in Ref.~\cite{ford1996},  which might cause some inconsistencies~\cite{fleming2011exact}. Using the mass renormalization and damping constants 
as defined in Ref.~\cite{ford1996}, one can see that the main modification from the exact integration is scaling of these quantities by the electron structure factor $f_{\omega_0}=\Omega^2/(\omega_0^2+\Omega^2)$. $f_{\omega_0}$ scaling factor here is important because although mass renormalization is obtained by infinite $\Omega$ limit, $\Omega$ needs to be finite for the definition of $\delta m$ to be meaningful and $f_{\omega_0}$ provides a cutoff for the calculated quantities at high oscillator frequency $\omega_0$, also. $I$ term in Eq.~(\ref{eq:t22}), which does not contribute to the master equation at the rotating wave approximation level, is also evaluated exactly here in terms of digamma functions and depends on the ratio of the temperature to the cutoff frequency of the blackbody field  as well as the oscillator frequency.

Using the exact integrals of $T_{1}^{(2)}$ and $T_{2}^{(2)}$ given by 
equations~(\ref{eq:t21}) and (\ref{eq:t22}) in Eq.~(\ref{eq:defAis}), we get the second order coefficients of master Eq.~(\ref{eq:secOrd})
\begin{eqnarray}
A_{1}^{(2)}&=&-f_{\omega_0}\frac{\delta m}{m},\label{eq:a1S}\quad
A_{2}^{(2)}=2f_{\omega_0}\frac{\gamma}{\omega_{0}} N\left(\beta\,\omega_{0}\right),\label{eq:a2S}\nonumber\\
A_{3}^{(2)}&=&f_{\omega_0}\frac{\gamma}{\omega_{0}}I,\label{eq:a3S}\quad
A_{4}^{(2)}=-f_{\omega_0}\frac{\gamma}{\omega_{0}}\label{eq:a4S},
\end{eqnarray}
\noindent where $N(\beta\,\omega_0)$ is the thermal occupation at inverse temperature $\beta$ and oscillator frequency $\omega_0$.

The third order contributions which are due, partly, to the diamagnetic term $A^2$ in the interaction Hamiltonian will be obtained by evaluating complex $T_{31}$ and $T_{32}$ and real $T_{33}$ and $T_{34}$ integrals defined in the Appendix including equations~(\ref{eq:t31})-(\ref{eq:t34}). These integrals can, also, be evaluated exactly by using the rational expansion of $\coth{(\beta\,\omega_{0})}$ function Eq.~(\ref{eq:coth}). The results are 
\begin{subequations}
\begin{eqnarray}
T_{31}&=&-\gamma f_{\omega_0}^2\left[\frac{\delta m}{m}
\coth{\left(\beta\,\omega_0\right)}-
\frac{\gamma}{\omega_0}\,I
\right]-i\,\omega_0\, f_{\omega_0}^2\left[\left(\frac{\gamma}{\omega_0}\right)^2-\left(\frac{\delta m}{m}\right)^2\right]\label{eq:ourT31},\\
T_{32}&=&-\gamma f_{\omega_0}^2\left[\frac{\gamma}{\omega_0}
\coth{\left(\beta\,\omega_0\right)}+
\frac{\delta m}{m}\,I
\right]+i\,2 \gamma\frac{\delta m}{m}f_{\omega_0}^2\label{eq:ourT32},\\
T_{33}&=&-\gamma f_{\omega_0}\frac{\delta m}{m}\left(\frac{1}{\beta\,\omega_0}+\coth{\left(\beta\,\omega_0\right)}\right)\label{eq:ourT33},\\
T_{34}&=&\frac{2}{\pi}\gamma f_{\omega_0}\frac{\delta m}{m}\Psi_{0}\left[\beta,\Omega,\omega_{0}\right]\label{eq:ourT34},
\end{eqnarray}
\end{subequations}
\noindent where $I$ is defined in Eq.~(\ref{eq:i1}), $\Psi_{0}\left[\beta,\Omega,\omega_{0}\right]=\psi_{0}\left[\frac{\beta\,\Omega}{\pi}\right]-\mathrm{Re}\{\psi_{0}\left[1-\frac{i\,\beta\,\omega_{0}}{\pi}\right]\}$ and $\psi_0(x)$ is the digamma function. The dominant term in all four integrals involve product of decay rate and the mass renormalization and their squares which are modulated by the cutoff factor $f_{\omega_0}$. The real parts of all four are temperature dependent while the imaginary parts of $T_{31}$ and $T_{32}$ depend on only the mass renormalization and the interaction strength terms. One should note that, if the cutoff frequency of the blackbody field is chosen as the causally allowed highest value as $\Omega=\tau_e^{-1}$ then $\frac{\delta m}{m}=1$. Noting that $\frac{\gamma}{\omega_0}\approx 10^{-7}$ for the optical frequencies, one can drop the $\Psi_0\left[\beta,\Omega,\omega_0\right]$ term of $T_{31}$ and $\coth{\left(\beta\omega_0\right)}$ term of $T_{32}$. 

The third order contribution to the coefficients of Eq.~(\ref{eq:secOrd}) can be obtained by plugging the results displayed in equations~(\ref{eq:ourT31})-(\ref{eq:ourT34}) into the definitions in Eq.~(\ref{eq:defAis}), which results
\begin{subequations}
\begin{eqnarray}
A_{1}^{(3)}&=&-f_{\omega_{0}}^{2}\,\left[\frac{\gamma^{2}}{ \omega_{0}^{2}}-\left(\frac{\delta m}{m}\right)^{2}\right],\label{eq:a13}\\
A_{2}^{(3)}&=&-f_{\omega_{0}}^{2}\,\frac{\gamma}{\omega_{0}}\left[\frac{\delta m}{m}\left(1+\frac{1}{f_{\omega_{0}}}\right)\left[1+2\,N\left(\omega_{0}\right)\right]-\frac{\gamma}{\omega_{0}}\,I\right]\nonumber\\
&&+f_{\omega_{0}}\frac{\gamma}{\omega_{0}}\,\frac{\delta m}{m}\left[2\,f_{\omega_{0}}-\frac{1}{\beta\,\omega_{0}}\right],\label{eq:a23}\\
A_{3}^{(3)}&=&-f_{\omega_{0}}^{2}\frac{\gamma}{\omega_{0}}\left[\frac{\gamma}{\omega_{0}}\left[1+2\,N\left(\beta\,\omega_{0}\right)\right]-\frac{\delta m}{m}\left(-I+\frac{2}{\pi}\frac{1}{f_{\omega_{0}}}\right)\right],\label{eq:a33}\\
A_{4}^{(3)}&=&2\,f_{\omega_{0}}^{2}\frac{\gamma}{\omega_{0}}\frac{\delta m}{m}\label{eq:a43}.
\end{eqnarray}
\end{subequations}
The most important difference between equations~(\ref{eq:a13})-(\ref{eq:a43}) and 
the third order coefficients in equation (7.16) of Ref.~\cite{ford1996} is the 
$f^2_{\omega_0}$ scaling which provides frequency cutoff for each one of those coefficients. We, also, provide exact expressions for $I_3$ and $I_4$ integrals of Ref.~\cite{ford1996} which were unevaluated there.

The fourth order contributions are tedious but straightforward to obtain 
by performing the integrals defined in Eq.~(\ref{eq:t46int}) and are given in equations~(\ref{eq:t41})-(\ref{eq:t63}) of the Appendix. Here, 
we collect the fourth-order contributions to the coefficients of the Eq.~(\ref{eq:secOrd}):
\begin{subequations}
\begin{eqnarray}
A_{1}^{(4)}&=&-f_{\omega_{0}}^{3}\,\left[\left(\frac{\gamma}{\omega_{0}}\right)^{2}\left(1-3\frac{\delta m}{m}\right)-\left(\frac{\gamma}{\Omega}\right)^{2}+\left(\frac{\delta m}{m}\right)^{3}\right],\label{eq:r14}\\
A_{2}^{(4)}&=&-f_{\omega_{0}}^{3}\,\frac{\gamma}{\omega_{0}}\left[\left(-\frac{5}{2}\,\frac{\gamma}{\Omega}+\frac{\delta m}{m}\left(\frac{1}{2}+\frac{\gamma}{\Omega}-3\,\frac{\delta m}{m}\right)\right)\left[1+2\,N\left(\beta\,\omega_{0}\right)\right]\right.\nonumber\\&&\left.+\frac{\gamma}{\omega_{0}}\,I
+2\,\frac{\gamma}{\Omega}+\frac{\delta m}{m}\left(3\frac{\delta m}{m}-\frac{\gamma}{\Omega}\right)\right]\nonumber\\
&&-f_{\omega_{0}}^{2}\,\gamma\,\beta\left[\frac{1}{\pi^{2}}\frac{\gamma}{\omega_{0}}\mathrm{Im}\left[\psi_{1}\left(\beta,\omega_{0}\right)\right]-\frac{1}{2}\frac{\delta m}{m}\,\csch^{2}{\left(\beta\,\omega_{0}\right)}\right],\\
A_{3}^{(4)}&=&-f_{\omega_{0}}^{3}\frac{\gamma}{\omega_{0}}\left[\frac{1}{2}\frac{\gamma}{\omega_{0}}\left(\left(\frac{\omega_{0}}{\Omega}\right)^{2}-1\right)\left[1+2\,N\left(\beta\,\omega_{0}\right)\right]-I\left(3\frac{\gamma}{\Omega}+\frac{\delta m}{m}\left(3\frac{\delta m}{m}-\frac{\gamma}{\Omega}\right)\right)\right.\nonumber\\
&&\left.+\frac{1}{\beta\,\Omega}\left(\frac{1}{f_{\omega_{0}}}\frac{\delta m}{m}\left(3\frac{\delta m}{m}+\frac{\gamma}{\Omega}\right)+\frac{\gamma}{\Omega}\left(1+\left(\frac{\omega_{0}}{\Omega}\right)^{2}\right)\right)\right]\nonumber\\
&&-f_{\omega_{0}}^{2}\,\gamma\,\beta\,\frac{\gamma}{\omega_{0}}\left[\frac{1}{2}\csch^{2}{\left(\beta\,\omega_{0}\right)}-\frac{2}{\pi^{2}}\left(1+2\,\frac{\Omega}{\gamma}\left(\frac{\delta m}{m}\right)^{2}\right)\psi_{1}\left[\frac{\beta\,\Omega}{\pi}\right]\right.\nonumber\\
&&\left.+\frac{1}{\pi^{2}}\frac{\omega_{0}}{\gamma}\left(\frac{\delta m}{m}\right)\left(\mathrm{Im}\{\psi_{1}\left(\beta,\omega_{0}\right)\}+\frac{1}{\pi}\frac{\beta\,\Omega}{f_{\omega_{0}}}\left(\frac{\Omega}{\omega_{0}}\right)\left(\frac{\delta m}{m}\right)\psi_{2}\left[\frac{\beta\,\Omega}{\pi}\right]\right)\right],\label{eq:r34}\\
A_{4}^{(4)}&=&-f_{\omega_{0}}^{3}\,\frac{\gamma}{\omega_{0}}\,\left[2\,\frac{\gamma}{\Omega}+\frac{\delta m}{m}\,\left(3\frac{\delta m}{m}-\frac{\gamma}{\Omega}\right)\right],\label{eq:r44}
\end{eqnarray}
\end{subequations}
\noindent where $\psi_{i}(x)$ is the polygamma function of order $i$ and $\mathrm{Im}\{\psi_{1}(\beta,\omega_0)\}=\mathrm{Im}\{\psi_1(1-i\beta\,\omega_0/\pi)\}$. All four coefficients have an overall $f^{3}_{\omega_0}$ scaling and are function of various combinations of interaction strength $\gamma/\omega_0$, 
mass renormalization $\delta m/m$ and $\gamma/\Omega$. $A_{1}^{(4)}$ and 
$A_{4}^{(4)}$ are independent of the temperature while $A_{2}^{(4)}$ and 
$A_{3}^{(4)}$ have complicated temperature dependence through thermal 
occupation factor as well as various polygamma functions.

\section{Discussion\label{discus}}

We will discuss the convergence of the expansion and the relative importance of various terms in Eq.~(\ref{eq:secOrd}) by a slight change of the master equation to 
\begin{eqnarray}
\frac{\partial \rho_{S}}{\partial t}&=&-\frac{1}{2\,m\,\hbar}\left\{
i\,\Delta\left[p^2,\rho_S\right]
+D_{xx}\left[p,\left[p,\rho_S\right]\right]\right.\nonumber \\
&&\left.+i\,m\,\omega_{0}\,\lambda \left[p,\left\{x,\rho_S\right\}\right]
+m\,\omega_{0}\,D_{xp}\left[p,\left[x,\rho_S\right]\right]\right\}
\label{eq:mast},
\end{eqnarray}
\noindent where $\Delta=\sum_{i}A^{(i)}_{1}$ is mass renormalization,  $\lambda=\sum_{i}A^{(i)}_{4}$ is decay constant, $D_{xx}=\sum_{i}
(A^{(i)}_{2}-A^{(i)}_{4})$ is normal diffusion and $D_{xp}=\sum_{i}A^{(i)}_{3}$ is anomalous diffusion coefficients. To discuss the relevance of various terms in Eq.~(\ref{eq:mast}), we will state the adjoint master equation that governs the dynamics of the moments of position and momentum operators of the oscillator as~\cite{breuer2002theory}
\begin{eqnarray}
\frac{d}{dt}\left\langle O\right\rangle_t&=&
\frac{i}{\hbar}\left\langle\left[H_{S},O\right]\right\rangle_t
+\frac{1}{2\,m\,\hbar}\{i\Delta\left\langle\left[p^2,O\right]\right\rangle_t
-D_{xx}\left\langle\left[p,\left[p,O\right]\right]\right\rangle_t\nonumber \\
&&+i\,m\,\omega_{0}\,\lambda \left\langle\left\{x,\left[p,O\right]\right\}\right\rangle_t
-m\,\omega_{0}\,D_{xp}\left\langle\left[p,\left[x,O\right]\right]\right\rangle_t\}
\label{eq:adjoint},
\end{eqnarray}
\noindent where $D_{xx}$ term causes diffusion of the variance of the 
oscillator position operator, $D_{xp}$ plays the same role for $x\cdot p$, 
$\Delta$ term is a change in the mass and $\lambda$ is decay constant of the position. It is obvious that while diffusion constants depend on the temperature, drift terms, mass renormalization and the damping constants, are temperature independent.  

By rearranging equations~(\ref{eq:a1S}), (\ref{eq:a13}) and (\ref{eq:r14}) the renormalization constant $\Delta$ can be expressed in terms of $\tau_e$, $\omega_0$ and $\Omega$ in a more suggestive form 
as
\begin{eqnarray}
\Delta=-\tilde{\Omega}\,f_{\tilde{\omega}_0}
\left\{
	1+\tilde{\Omega}\,f_{\tilde{\omega}_0}
	\left[
    	\left(
        	\tilde{\omega}_0^2-1
        \right)+f_{\tilde{\omega}_0}
        \left(
        	\tilde{\Omega}\,
            \left(1-3\tilde{\omega}_0^2\right)
            +\tilde{\omega}_0^2 \left(1-\tilde{\omega}_0^2\right)
      	\right)
  	\right]
\right\}
\label{eq:delta},
\end{eqnarray}
\noindent where $\tilde{\omega}_0=\omega_0/\Omega$ and $\tilde{\Omega}=\Omega\,\tau_e$ are the dimensionless oscillator and the bath cutoff frequency, respectively and $f_{\tilde{\omega}_0}=1/(1+\tilde{\omega}_0^2)$ is the bath cutoff function. 
Note that the inverse of $\tilde{\omega}_0$ can be considered as a resonance factor, its a measure 
of the effectiveness of the interaction between the charged oscillator and the 
environment, a large or small value of $\tilde{\omega}_0$ indicates that the oscillator is 
detuned from the peak of the environmental spectral distribution. The 
successive terms in the nested parenthesis of the expression for $\Delta$ 
come from the second, the third and fourth order corrections, respectively. 
These terms are in the form of product of mass renormalization and cutoff 
factor modulated with the ratio of the oscillator frequency to the bath 
cutoff. The ratio of the third  and the fourth order to the second order 
contribution can be expressed as $\tilde{\Omega}(\tilde{\omega}_0^2-1)/(\tilde{\omega}_0^2+1)$ and $\tilde{\Omega}^2 
(1-3\tilde{\omega}_0^2)/(\tilde{\omega}_0^2+1)^2$ which approach $\tilde{\Omega}$ and $0$ at 
 the high frequency limit $\omega_0\gg\Omega$, respectively. In the low 
 frequency limit $\tilde{\omega}_0\ll 1$, these ratios tend to $-\tilde{\Omega}$ and $\tilde{\Omega}^2$.
 
 Similar to $\Delta$, we can rewrite $\lambda$ in a simpler form as
\begin{equation}
\lambda=-f_{\tilde{\omega}_0}\,\tilde{\omega}_0\,\tilde{\Omega}\,\left\{1+f_{\tilde{\omega}_0}\tilde{\Omega}
\left[-2+f_{\tilde{\omega}_0}\left(\tilde{\Omega}(3-\tilde{\omega}_0^2)+2\tilde{\omega}_0^2\right)\right]\right\}
\label{eq:lambda}.
\end{equation}
\noindent The perturbative terms enter into the expression of $\lambda$ as powers of $f_{\tilde{\omega}_0}\tilde{\Omega}$. A comparison of equations~(\ref{eq:delta}) and (\ref{eq:lambda}) shows that the leading terms of $\Delta$ and $\lambda$ are $f_{\tilde{\omega}_0}\,\tilde{\Omega}$ and $f_{\tilde{\omega}_0}\,\tilde{\omega}_0\,\tilde{\Omega}$, respectively. So, when the oscillator is in resonance with 
the peak frequency of the environment, the magnitude of renormalization and the damping constants would be comparable while off-resonance $\lambda$ is expected to be small compared to $\Delta$ for low frequencies. At high frequencies, $f_{\tilde{\omega}_0}\rightarrow 0$ and $\Delta$ and $\lambda$ tend to zero.

\begin{figure}[!ht]
   \centering
   \includegraphics[width=14cm]{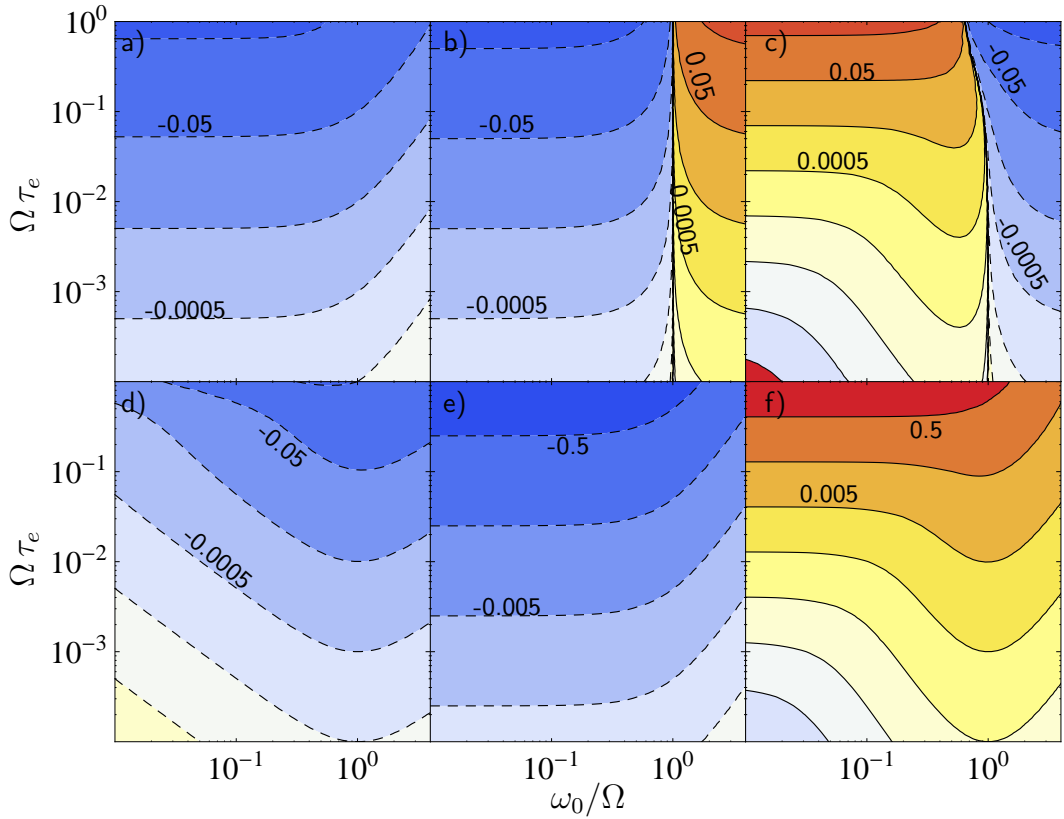}
   \caption{Magnitude (a) and ratio of the third (b) and fourth  order contributions (c) to the second order term for $\Delta$ and $\gamma$ as function of the oscillator frequency $\omega_0$ and the cutoff frequency of the blackbody environment $\Omega$. Both axis scales and the contour separations are logarithmic and the dashed lines indicate negative contours.}
   \label{fig:a2a3}
\end{figure}

\begin{figure}[!ht]
   \centering
   \includegraphics[width=14cm]{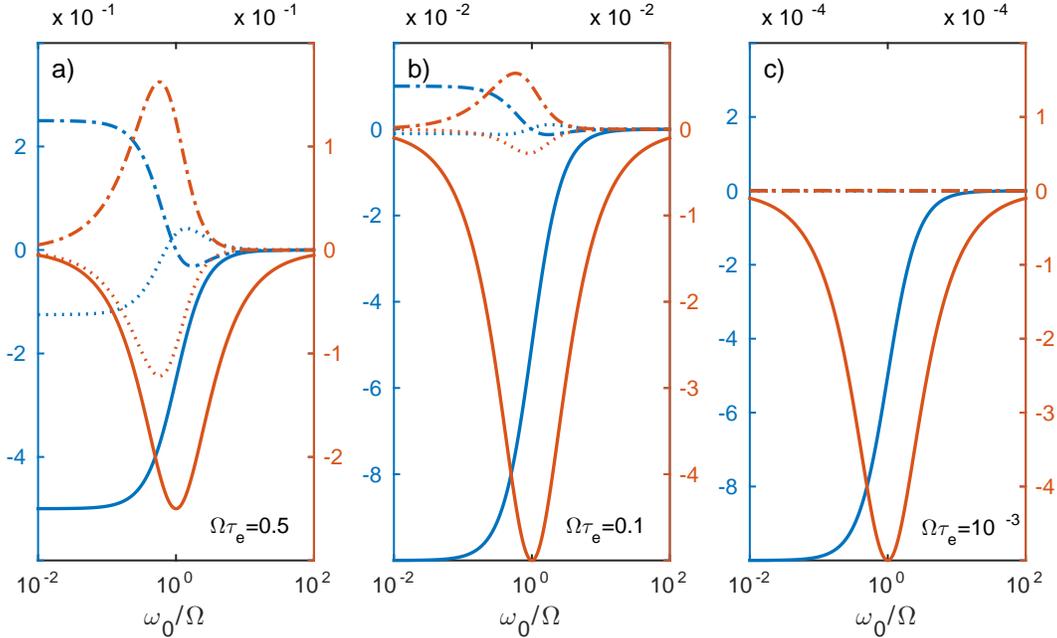}
   \caption{The second (straight), the third (dot-dashed) and the fourth (dotted) order contributions to mass renormalization (left axis) and damping rate (right axis) as function of the dimensionless oscillator frequency at three different 
   environmental cutoff frequencies (a) $\Omega\,\tau_e=0.5$, b) $\Omega\,\tau_e=0.1$ and c) $\Omega\,\tau_e=0.001$).}
   \label{fig:dg}
\end{figure}

The magnitude of $\Delta$ and $\lambda$ along with the ratio of the third and the fourth order contributions to their leading term, as function of $\Omega\,\tau_e$ and $\omega_0/\Omega$, are presented in Fig.~\ref{fig:a2a3}(a)-(c) and Fig.~\ref{fig:a2a3}(d)-(f), respectively. Both the renormalization and the decay constant are negative for the considered frequency range. $\Delta^{(3)}/\Delta$ and $\Delta^{(4)}/\Delta$ changes sign at $\omega_0=\Omega$ while $\lambda^{(3)}/\lambda<0$ and $\lambda^{(4)}/\lambda>0$ for the whole range of $\omega_0/\Omega$ and $\Omega$ considered here. These ratios can be considered as a proxy for the convergence of the perturbative expansion of the master equation. It is clear that $B^{(3)}$ and $B^{(4)}$ contributions partially cancel each other and only the second order term $B^{(2)}$ might be sufficient to account for the dynamics of the oscillator for the most part of the considered parameter range. Only very near the upper limit of the cutoff frequency ($\Omega\approx \tau_e^{-1}$), the whole perturbative treatment might be divergent because the magnitude as well as the contributions approach $-1$ and $\pm 1$, respectively. 
It is probable that including even higher order terms in the master equation will not lead to a converged $\Delta$ and $\lambda$ in this parameter regime. 
One should note that the interaction strength in the current model is 
$\gamma/\omega_0=\omega_0\tau_e$ and the non-convergence issue is independent of its value for high cutoff $\Omega$. We display the $\omega_0$ dependent $\Delta$ and $\lambda$ at three different 
$\Omega\,\tau_e$ values in Fig.~(\ref{fig:dg}) which indicates that while $B^{(3)}$ and $B^{(4)}$ contributions are significant at $\Omega\,\tau_e=0.5$, they decrease substantially when $\Omega\,\tau_e$ is reduced to $0.1$ and become insignificant at $\Omega\,\tau_e=10^{-3}$.
It can be deduced from Fig.~\ref{fig:a2a3}(b), (c), (e) and (f) that the relative contributions of the third and the fourth order terms are less than  $1\,\%$ for $\Omega\,\tau_e<0.01$ and including only the second order terms in master equation might be sufficient for an accurate description of the dynamics of the charged oscillator.

Both the normal and the anomalous diffusion coefficients are temperature dependent. We will first discuss the high temperature limit of the higher 
order contributions to $D_{xx}$ and $D_{xp}$. Defining $z=\beta\,\Omega$, $z\ll 1$ corresponds to the 
high temperature regime and Taylor expanding hyperbolic cotangent and polygamma functions around zero and taking the first term of the expansion as, $\coth{\left(\tilde{\omega}_0\,z\right)}\rightarrow 1/(\tilde{\omega}_0\,z)$, $\csch{\tilde{\omega}_0\,z}\rightarrow 1/\tilde{\omega}_0\,z$, $\psi_{1}(1\pm\,i\,\tilde{\omega}_0\,z/\pi)\rightarrow \pi^2/6$, $\psi_{1}(z/\pi)\rightarrow  \pi^2/z^2$ and $\psi_{2}(z/\pi)\rightarrow  -2 \pi^3/z^3$
we get
\begin{eqnarray}
D_{xx}^{(2)}&=&f_{\tilde{\omega}_0}\,\frac{\tilde{\Omega}}{z} \label{eq:dxxHighT1}\nonumber,\\
D_{xx}^{(3)}&=&-f^2_{\tilde{\omega}_0}\,\frac{\tilde{\Omega}^2}{z}\left(3+\tilde{\omega}_0^2\right)\label{eq:dxxHighT2},\\
D_{xx}^{(4)}&=&f^3_{\tilde{\omega}_0}\,\frac{\tilde{\Omega}^2}{z}\left[3\tilde{\Omega}+\tilde{\omega}_0^2 (2-\tilde{\Omega})\right]
\label{eq:dxxHighT3},\nonumber
\end{eqnarray}
\noindent and
\begin{eqnarray}
D_{xp}^{(2)}&=&f_{\tilde{\omega}_0}\,\tilde{\omega}_0\,\frac{\tilde{\Omega}}{z}\label{eq:dxpHighT1}\nonumber,\\
D_{xp}^{(3)}&=&-2f^2_{\tilde{\omega}_0}\,\tilde{\omega}_0\,\frac{\tilde{\Omega}^2}{z}\left(2+\tilde{\omega}_0^2\right)\label{eq:dxpHighT2},\\
D_{xp}^{(4)}&=&f^3_{\tilde{\omega}_0}\,\tilde{\omega}_0\,\frac{\tilde{\Omega}^2}{z}\left[6\tilde{\Omega}+\tilde{\omega}_0^2(3+\tilde{\omega}_0^2)(1+\tilde{\Omega})\right],
\label{eq:dxpHighT3}\nonumber
\end{eqnarray}
which indicate that all contributions to $D_{xx}$ and $D_{xp}$ are linear in temperature 
at the high temperature limit. Both the second and the fourth order contributions to both diffusion coefficients are positive while the third order ones are always negative.

One should note that from a comparison of equations~(\ref{eq:dxxHighT2})-(\ref{eq:dxpHighT2}) that the terms contributing to $D_{xp}$ contain an $\tilde{\omega}_0$ factor which is a measure of interaction strength. So, anomalous diffusion coefficient is expected to be much smaller than the normal diffusion coefficient in the weak coupling limit. 
In this limit ($\tilde{\omega}_0\rightarrow 0$), $f(\tilde{\omega}_0)=1$ and the convergence of the coefficients depends solely on the cutoff frequency of the environment as can be seen from equations~(\ref{eq:dxxHighT2})-(\ref{eq:dxpHighT2}).     

For the low temperature limit $z\gg 1$, we can use the approximations 
$\coth{(z)}\rightarrow 1$, 
$\csch{(z)}\rightarrow 0$, $\textrm{Re}\{\psi_0(1-i\,\tilde{\omega}_0\,z/\pi)-\psi_0(z/\pi)\}\rightarrow\log{(\tilde{\omega}_0)}$,
$z\,\textrm{Im}\left[\psi_1(1-i\,\tilde{\omega}_0\,z/\pi)\right]\rightarrow \pi/\tilde{\omega}_0$, $z^2 \psi_{2}(z/\pi)\rightarrow -\pi^2$ and $z\psi_1(z/\pi)\rightarrow \pi$ to express $D_{xx}$ and $D_{xp}$ as 
\begin{eqnarray}
D_{xx}^{(2)}=&&f_{\tilde{\omega}_0}\,\tilde{\omega}_0\,\tilde{\Omega},\nonumber\\
D_{xx}^{(3)}=&&-f_{\tilde{\omega}_0}^2\,\tilde{\omega}_0\,\tilde{\Omega}^2\,\left[2+\tilde{\omega}_0^2-\frac{2}{\pi}\,\tilde{\omega}_0\,\log{(\tilde{\omega}_0)}\right],\\
D_{xx}^{(4)}=&&-f_{\tilde{\omega}_0}^3\,\tilde{\omega}_0\,\tilde{\Omega}^2\,\left\{\frac{1}{2}-3\tilde{\Omega}+(\tilde{\Omega}-\frac{5}{2})\,\tilde{\omega}_0^2+\frac{\tilde{\omega}_0}{\pi}\left[1+\tilde{\omega}_0^2+2\log{(\tilde{\omega}_0)}\right]\right\},\nonumber
\end{eqnarray} and
\begin{eqnarray}
D_{xp}^{(2)}=&&\frac{2}{\pi}f_{\tilde{\omega}_0}\,\tilde{\omega}_0\,\tilde{\Omega}\log{(\tilde{\omega}_0)},\nonumber\\
D_{xp}^{(3)}=&&-f_{\tilde{\omega}_0}^2\,\tilde{\omega}_0\,\tilde{\Omega}^2\left[\tilde{\omega}_0+
\frac{2}{\pi}(2+\tilde{\omega}_0^2)\log{(\tilde{\omega}_0)}\right],\\
D_{xp}^{(4)}=&&\frac{1}{\pi}f_{\tilde{\omega}_0}^3\,\tilde{\omega}_0\,\tilde{\Omega}^2\,\left\{
5\tilde{\Omega}-1+\tilde{\omega}_0^4(2+\tilde{\Omega})+(1+6\tilde{\Omega})\tilde{\omega}_0^2+2\left[3\tilde{\omega}_0^2+\tilde{\Omega}(3-\tilde{\omega}_0^2)\right]\log{(\tilde{\omega}_0)}\right.\nonumber\\
&&+\left.\frac{\pi}{2}\tilde{\omega}_0(1-\tilde{\omega}_0^2)\right\}.\nonumber
\end{eqnarray}
\noindent The $\omega_{0}/\Omega$ dependence of $D_{xx}$ and $D_{xp}$ should be the qualitatively different for the low and high environment temperatures based on the spectral distribution, $I(\omega_{0})=J(\omega_{0})\coth{\left(\beta\,\omega_0\right)}$, of the environment.

\begin{figure}[!ht]
   \centering
   \includegraphics[width=14cm]{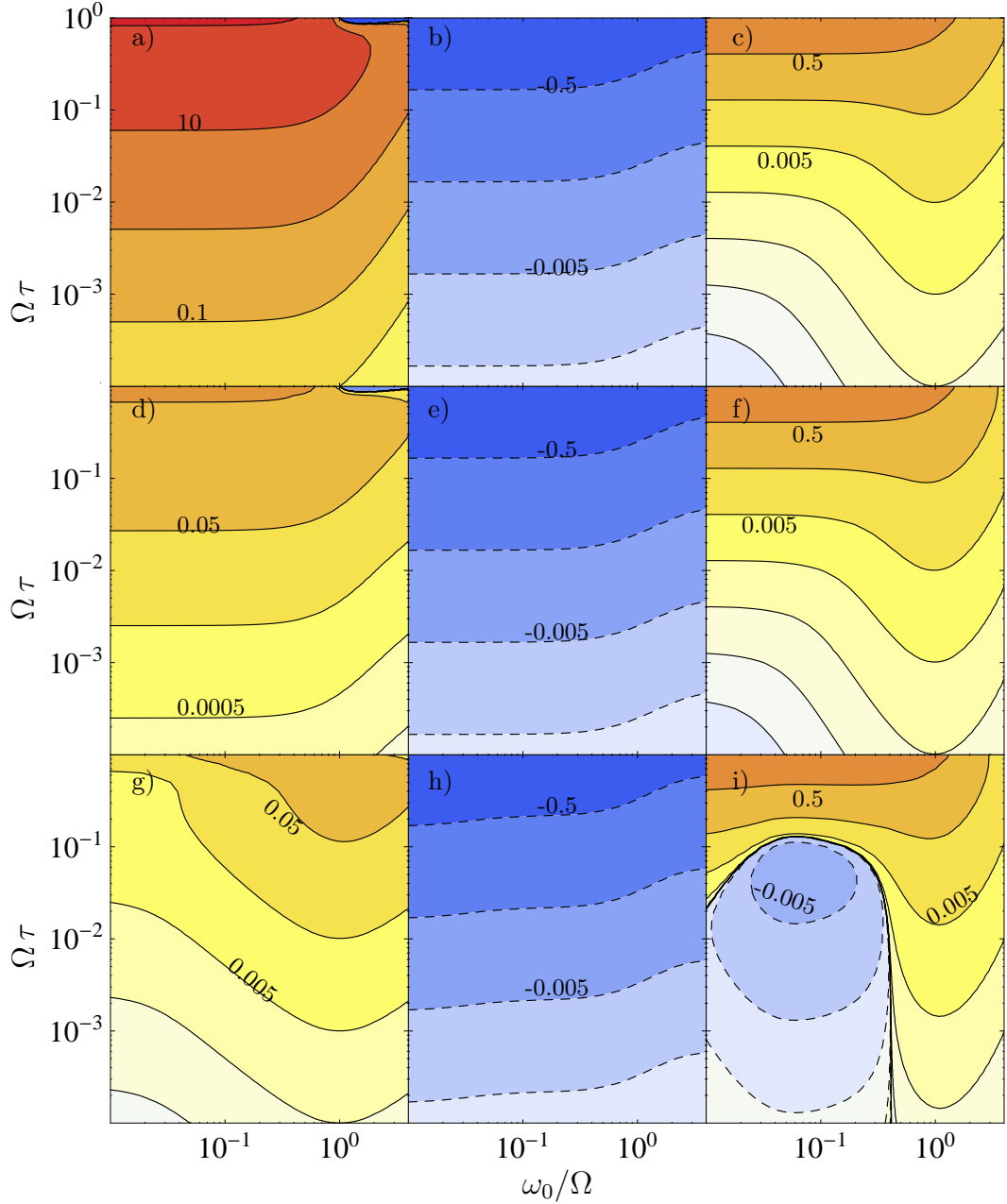}
   \caption{Magnitude (a, d, g) and the ratio of the third (b, e, h) and the fourth order contributions (c, f, i) to the second order term for the normal diffusion coefficient $D_{xx}$  as function of the oscillator frequency $\omega_0$ and the cutoff frequency of the black-body environment $\Omega$ at high ($k_B T=0.01\,\hbar\,\Omega$) (a, b, c), intermediate ($k_B T=\hbar\,\Omega$) (d, e, f) and low ($k_B T=100\,\hbar\,\Omega$) (g, h, i) temperatures.}
   \label{fig:dxx}
\end{figure}

We have considered low ($k_B T=100\,\hbar\,\Omega$), intermediate ($k_B T= \hbar \Omega$) and high ($k_B T=0.01\,\hbar\,\Omega$) temperature limits of the environment and display the magnitude and the ratio of the third and the fourth 
order contributions to the second order one for the normal diffusion coefficient $D_{xx}$ in Fig.~\ref{fig:dxx}(a)-(i). From the first column of the figure, it is obvious that $D_{xx}$ increases with increasing initial temperature as expected, because of the increase in the thermal excitations with increasing temperature. Similar to the drift terms, the third and fourth order contributions to $D_{xx}$ have opposite signs for intermediate and high $T$ values while at low temperatures $D^{(4)}_{xx}$ is negative or positive depending on $\omega_0$ and $\Omega$. 
The interaction strength dependence of $D_{xx}$ 
changes somewhat from high to low temperature region (compare Fig.~\ref{fig:dxx}(a) and Fig.~\ref{fig:dxx}(g); while it is almost independent 
of $\tilde{\omega}_0=\omega_0/\Omega\approx 1$ at high temperatures, it increases with 
decreasing $\tilde{\omega}_0$ in the low temperatures. Fig.~\ref{fig:dxx}(b), (e) and (h) indicate that the ratio $D_{xx}^{(3)}/D_{xx}^{(2)}$ is always negative and 
almost independent of the temperature. Similarly, 
$D_{xx}^{(4)}/D_{xx}^{(2)}$ (Fig.~\ref{fig:dxx}(c), (f) and (i)) is positive 
and almost identical at high and intermediate temperatures while it becomes negative at low $\omega_0$ and $\Omega$ region at low temperatures. Since $D_{xx}^{(3)}$ and $D_{xx}^{(4)}$ have opposite signs for the most of the 
considered parameter range, instead of taking into account only the second and third order contributions, keeping only the second order terms might be more accurate.

\begin{figure}[!ht]
   \centering
   \includegraphics[width=14cm]{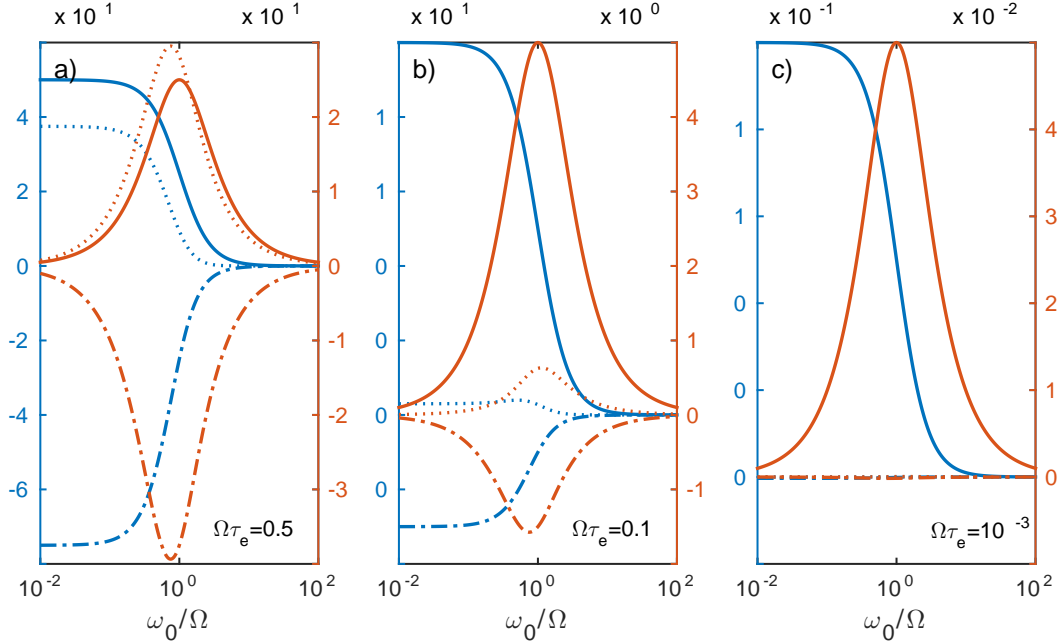}
   \caption{Magnitude of $B^{(2)}$ (straight), $B^{(3)}$ (dot-dashed) and $B^{(4)}$ (dotted) contributions to normal ($D_{xx}$: left axis) and anomalous diffusion coefficients ($D_{xp}$: right axis) at high temperature ($\beta\,\Omega=0.01$) as function of the oscillator frequency at a) $\Omega\,\tau_e=0.5$, b) $\Omega\,\tau_e=0.1$  and c) $\Omega\,\tau_e=10^{-3}$.}
   \label{fig:htdxp}
\end{figure}

\begin{figure}[!ht]
   \centering
   \includegraphics[width=14cm]{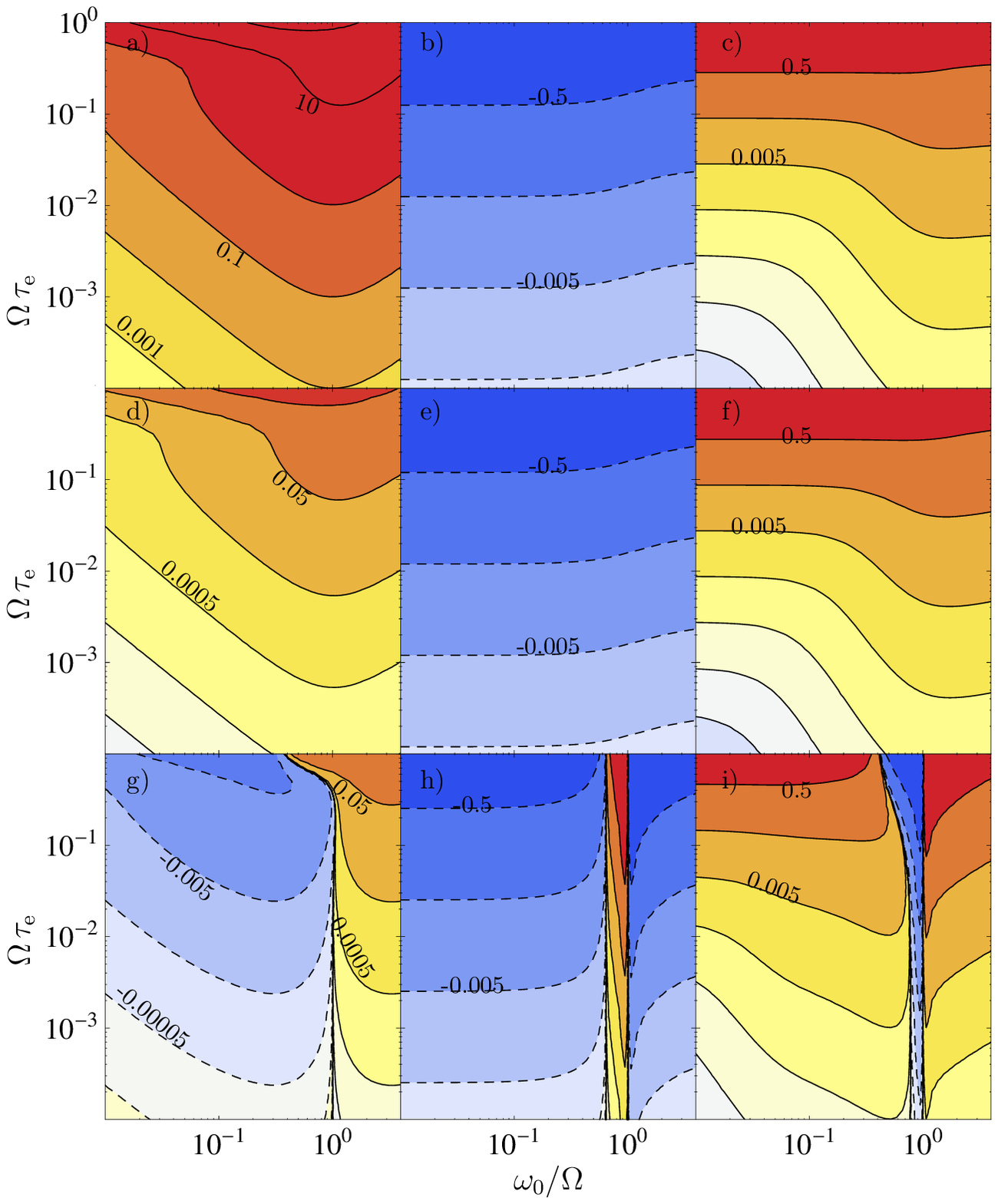}
   \caption{Same as figure~\ref{fig:dxx} for $D_{xp}$.}
   \label{fig:dxp}
\end{figure}

\begin{figure}[!ht]
   \centering
   \includegraphics[width=14cm]{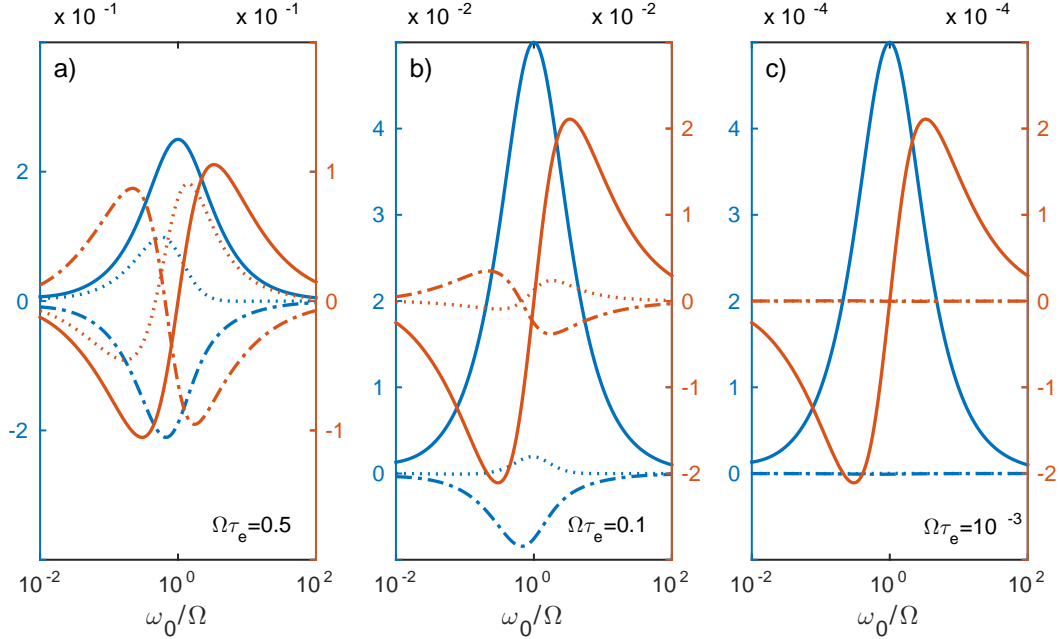}
   \caption{Same as figure~\ref{fig:htdxp} for low temperature ($\beta\,\Omega=100$).}
   \label{fig:ltdxp}
\end{figure}

To delineate the relative importance of higher order terms for the anomalous diffusion coefficient $D_{xp}$, we display its value and the relative third and the fourth order contributions as function of $\omega_0$ and $\Omega$ in Fig.~\ref{fig:dxp}(a)-(i) at low, intermediate and high temperatures. 
The magnitude of $D_{xp}$ originating from $B^{(2)}$, $B^{(3)}$ and $B^{(4)}$ are, also, displayed 
in Fig.~(\ref{fig:htdxp}) and Fig.~(\ref{fig:ltdxp}) for different $\Omega$ values at high and low temperatures, respectively.

Although $D_{xp}$ at high and intermediate temperatures show similar behavior as the normal diffusion coefficient (compare the first two rows of Fig.~(\ref{fig:dxx}) and Fig.~(\ref{fig:dxp}), such as being positive for all $\omega_0$ and $\Omega$ values and $D_{xp}^{(3)}$ and $D_{xp}^{(4)}$ having opposite signs, it becomes negative for $\tilde{\omega}_0<1$ at low temperatures (see Fig.~\ref{fig:dxp}(g)). Sign of $D_{xp}$ is related with the squeezing of the position 
or the momentum of the oscillator~\cite{paz2009}. The changing sign indicates a change in localized dynamical variable depending on the initial temperature of the environment being high or low.

The qualitative difference in both $D_{xx}$ and $D_{xp}$ and the contributions to them in all the
considered orders between the high and low temperature environments is prominent when one compares 
Fig.~(\ref{fig:htdxp}) and Fig.~(\ref{fig:ltdxp}). Independent of the temperature and the bath cutoff 
frequency, the shapes of the absolute values of the second, the third and the fourth order diffusion 
terms as function of the oscillator frequency are similar: the components of $D_{xp}$ ($D_{xx}$) at high (low) temperatures display a single resonance peak at $\omega_0=\Omega$, while the low
temperature behavior of $D_{xp}$ is similar to an anti-resonance structure. Such behavior can be understood by referring to the spectral density function.

\section{Conclusions\label{conc}}
We have considered the exact treatment of the third and the fourth order corrections to the 
master equation of a charged harmonic oscillator interacting with a black body radiation field by using Krylov averaging method. The oscillator-bath interaction is assumed to be in velocity-coupling form. Diamagnetic interaction term, which is generally neglected, is also taken into account. It is found that including higher than the second order terms would not change the structure of the master equation, \emph{i.e.} they do not introduce any new commutators. 

For mass renormalization, decay constant, normal, and anomalous diffusion coefficients for the oscillator in the electromagnetic field, we obtained exact analytical expressions, to the fourth order, which indicate that 
mass renormalization and decay constant depend on the oscillator frequency and the bath cutoff frequency while the diffusion coefficients are sensitive to the bath temperature along with those two variables. Anomalous diffusion coefficient is found to go under a qualitative change between the 
high and the low temperature limits of the environment.

The aim of this study was to delineate the conditions under which the higher order corrections to the master equation might make a difference in the dynamics of the charged harmonic oscillator in electromagnetic field. It is found that the single most important parameter in that context is the 
chosen cutoff frequency $\Omega$ compared to the inverse of electron time $\tau_e$. 
We have shown that the second and the third order contributions are, mostly, opposite in sign and cancel each other. So, keeping only the second order terms in the master equation would be accurate enough for the weak interaction regime and $\Omega\,\tau_{e}<0.01$. For $\Omega\,\tau_{e}>0.1$, a perturbative expansion would not be appropriate as the order of consecutive terms have comparable magnitudes.

\newpage
\appendix
\section{Krylov Method of Averaging for Deriving Master Equation}
Here, we summarize the Krylov averaging method used in Ref.~\cite{ford1996}, obtain exact analytical 
expressions for the integrals involved in the definition of $B^{(2)}\left(\rho_{S},t\right)$ and $B^{(3)}\left(\rho_{S},t\right)$ and give a 
detailed derivation of the fourth order correction $B^{(4)}\left(\rho_{S},t\right)$.

Since the system-bath interaction is assumed to be weak, one would expect that the change in the product form of the initial density matrix of the total system is minimal and $\rho(t)$ can be expressed as $\rho(t)\cong \rho_S\otimes\rho_R$ where the difference from the product terms 
can be approximated as a power expansion in the perturbation parameter $\epsilon$ and a function of the oscillator density matrix $\rho_S$ as:
\begin{equation}
\label{eq:genForm}
\rho\left(t\right)=\rho_{S}\left(t\right)\otimes\rho_{R}+\epsilon\, F^{(1)}\left(\rho_{S},t\right)+\epsilon^{2}\,F^{\left(2\right)}\left(\rho_{S},t\right)+\dots,
\end{equation}
\noindent where terms containing $\epsilon$ represent small fluctuations in the amplitude of the density matrix of the whole system around $\rho_{S}\left(t\right)\otimes\rho_{R}$.
Similarly, the time rate of change of the system density matrix can be written as 
\begin{equation}
\label{eq:TimeDepDen}
\frac{\partial \rho_{S}\left(t\right)}{\partial t}=\epsilon\, B^{(1)}\left(\rho_{S},t\right)+\epsilon^{2}\,B^{(2)}\left(\rho_{S},t\right)+\dots\,\,.
\end{equation}
\noindent  For the definition $\rho_S(t)=\mathrm{Tr}_R\{\rho(t)\}$ to hold, $\mathrm{Tr}_{R}\{F^{(n)}\left(\tilde{\sigma},t\right)\}=0$ for $n=1,2,\dots$\\. Also, $B^{(n)}\left(\rho_{S},t\right)$ should be independent of bath operators, so $\mathrm{Tr}_{R}\{B^{\left(n\right)}\left(\rho_{S},t\right)\rho_{R}\}=B^{(n)}\left(\rho_{S},t\right)$.

To obtain $B^{(n)}\left(\rho_{S},t\right)$s, we insert $\rho(t)$ of Eq.~(\ref{eq:genForm}) into the von Neumann equation (Eq.~(\ref{eq:IntNeumann})) and use Eq.~(\ref{eq:TimeDepDen}) for the time dependence of the reduced density matrix of the system and match the coefficients of the equal powers of $\epsilon$ on the left and the right sides. The process produces a hierarchy of 
coupled equations for $B^{(n)}\left(\rho_{S},t\right)$ and $F^{(n)}\left(\rho_{S},t\right)$ as:
\begin{eqnarray}
\label{eq:gen}
\rho_{R}\otimes B^{(1)}+\frac{\partial F^{\left(1\right)}}{\partial t}&=&\frac{1}{i\,\hbar}\left[H_{I}\left(t\right),\rho_{S}\otimes\rho_{R}\right]\nonumber \\
\rho_{R}\otimes B^{\left(n\right)}+\frac{\partial F^{\left(n\right)}}{\partial t}&=&\frac{1}{i\,\hbar}\left[H_{I}\left(t\right),F^{\left(n-1\right)}\right]-\sum\limits_{m=1}^{n-1}F^{\left(n-m\right)}\left(B^{\left(m\right)},t\right).
\end{eqnarray}
The set of equations in (\ref{eq:gen}) can be solved by starting from $B^{(1)}\left(\rho_{S},t\right)$ and $F^{(1)}\left(\rho_{S},t\right)$ and invoking the trace over the bath conditions mentioned above. We quote the results, up to order four, from Ref.~\cite{ford1996}:
\begin{subequations}
\begin{eqnarray}
\label{eq:group}
B^{\left(1\right)}\left(\rho_{S},t_{1}\right)&=&\frac{1}{i\,\hbar}\{1\},\label{eq:b1a}\\
B^{\left(2\right)}\left(\rho_{S},t_{1}\right)&=&\frac{1}{\left(i\,\hbar\right)^{2}}\int_{-\infty}^{t_{1}}dt_{2}\left(\{12\}-\{1:2\}\right),\label{eq:b2b}\\
B^{\left(3\right)}\left(\rho_{S},t_{1}\right)&=&\frac{1}{\left(i\,\hbar\right)^3}\int_{-\infty}^{t_{1}}dt_{2}\int_{-\infty}^{t_{2}}dt_{3}\left\{\{123\}-\{12:3\}-\{13:2\}-\{1:23\}\right.\nonumber\\
&&+\left.\{1:2:3\}+\{1:3:2\}\right\},\label{eq:b3c}\\
B^{\left(4\right)}\left(\rho_{S},t_{1}\right)&=&\frac{1}{\left(i\,\hbar\right)^4}\int_{-\infty}^{t_{1}}dt_{2}\int_{-\infty}^{t_{2}}dt_{3}\int_{-\infty}^{t_{3}}dt_{4}\left\{\{1234\}-\{12:34\}-\{13:24\}\right.\nonumber\\
&&-\left.\{14:23\}-\{1:234\}-\{123:4\}-\{124:3\}-\{134:2\}\right.\nonumber\\
&&+\left.\{12:3:4\}+\{12:4:3\}+\{13:2:4\}+\{13:4:2\}\right.\nonumber\\
&&+\left.\{14:3:2\}+\{1:2:34\}+\{1:3:24\}+\{1:4:23\}\right.\nonumber\\
&&+\left.\{1:24:3\}+\{1:34:2\}+\{14:2:3\}+\{1:23:4\}\right.\nonumber\\
&&-\left.\{1:3:2:4\}-\{1:3:4:2\}-\{1:4:2:3\}\right.\nonumber\\
&&-\left.\{1:2:3:4\}-\{1:2:4:3\}-\{1:4:3:2\}\right\},\label{eq:b4d}
\end{eqnarray}
\end{subequations}
\noindent where the braces $\{\dots\}$ represent the trace over the reservoir degrees of freedom of the nested commutators of the closed-system Hamiltonian at times $t_{i}$ with the density matrix $\rho_{S}\otimes\rho_R$ while the column sign indicates a further trace operation:
\begin{eqnarray*}
\{1234\}&=&\mathrm{Tr}_{R}\{\left[H_{I}\left(t_{1}\right),\left[H_{I}\left(t_{2}\right),\left[H_{I}\left(t_{3}\right),\left[H_{I}\left(t_{4}\right),\rho_{S}\,\rho_{R}\right]\right]\right]\right]\}\\
\{12:34\}&=&\mathrm{Tr}_{R}\{\left[H_{I}\left(t_{1}\right),\left[H_{I}\left(t_{2}\right),Tr_{R}\{\left[H_{I}\left(t_{3}\right),\left[H_{I}\left(t_{4}\right),\rho_{S}\,\rho_{R}\right]\right]\}\rho_{R}\right]\right]\}.
\end{eqnarray*}
Thermal expectation value of the products of field operators at different times will be represented by using $\langle ij\ldots \rangle$ notation as 
\begin{eqnarray*}
\langle 1 2 3\ldots\rangle &\equiv& \mathrm{Tr}_{R}\{\rho_{R}\,A\left(t_{1}\right)A\left(t_{2}\right)A\left(t_{3}\right)\ldots\}.\label{traceR}
\end{eqnarray*}
Since $A\left(t\right)$ is a Gaussian operator, the thermal expectation value of the product terms that contain an odd number of field operators are zero while those that contain an even number of $A(t_i)$ can be expressed as the sum of product of pair expectations by using Isserlis' theorem: 
\begin{eqnarray}
\langle 1\rangle &=&\langle 123\rangle=0\label{odd},\nonumber\\
\langle1234\rangle &=&\langle 12\rangle \langle 34\rangle+\langle 13\rangle \langle 24\rangle+\langle 14\rangle \langle 23\rangle\label{even}.
\end{eqnarray}
 Explicit form of the pair expectation $\langle12\rangle$ for the present problem can be easily calculated by using the time dependent field operator $A(t_i)$ as defined in Eq.~(\ref{eq:Arfunc})~\cite{schlosshauer2007decoherence}:
 \begin{equation}
 \label{eq:exp}
\langle 12\rangle=\frac{\hbar}{\pi}\int_{0}^{\infty}d\omega\, J\left(\omega\right)
\{\coth\left(\frac{\hbar\,\omega}{2\,k_{B}\,T}\right)\cos\left[\omega\left(t_{1}-t_{2}\right)\right]
-i\,\sin\left[\omega\left(t_{1}-t_{2}\right)\right]\},
\end{equation}
where $J\left(\omega\right)$ denotes the spectral density of the bath and contains all the information related to the bath-oscillator interaction and is given in Eq.~(\ref{spectralDensity}) for the present problem.

Before evaluating $B^{\left(n\right)}\left(\rho_{S},t_{1}\right)$ for $n=1,2,3,4$ of equations~(\ref{eq:b1a})-(\ref{eq:b4d}), we should note that contribution of $B^{(n)}$s to the dynamics of the oscillator density matrix has the same form of independent of $n$, as:
\begin{eqnarray}
\label{eq:CofB3}
\frac{\partial \rho_{S}}{\partial t}&=&-\frac{1}{2\,\hbar\, m\,\omega_{0}}\left(\left[p\left(t\right),\{T_1\,p\left(t\right)+m\,\omega_{0}\,T_2\,x\left(t\right)\right\}\rho_{S}]\right.\nonumber\\
&&+\left.\left[\rho_{S}\{T^{*}_1\,p\left(t\right)+m\,\omega_{0}\,T^{*}_2\,x\left(t\right)\},p\left(t\right)\right]\right),
\end{eqnarray}
\noindent where $T_1$ and $T_2$ are time integrals of various combinations 
of pair correlation functions and can be written as $T_{i}=\sum_{n}T_{i}^{(n)}$ for corrections up to order $n$. We will provide the details of calculating $T_{i}^{(n)}$ up to $n=4$. 

In the remainder of the Appendix, we will outline the evaluation of the contributions to the coefficients of the commutators in Eq.~(\ref{eq:CofB3}) up to fourth order. First, we note that $B^{\left(1\right)}\left(\rho_{S},t_{1}\right)=0$ because of the Gaussian property of $A(t)$. The second order contribution is, also, easy to calculate: From Eq.~(\ref{eq:b2b}), the $\{1:2\}$ term is zero and $\{12\}$ term gives $T_{1}^{(2)}=T_{21}$ and $T_{2}^{(2)}=T_{22}$ as: 
\begin{subequations}
\begin{eqnarray}
T_{21}&=&\left(\frac{2\,\omega_{0}}{\hbar\, m}\right)\int_{0}^{\infty}dt_{12}\;\langle 12\rangle \cos\left(\omega_{0}\,t_{12}\right)\label{T21},\\
T_{22}&=&\left(\frac{2\,\omega_{0}}{\hbar\, m}\right)\int_{0}^{\infty}dt_{12}\;\langle 12 \rangle\sin\left(\omega_{0}\,t_{12}\right)\label{T22}.
\end{eqnarray}
\end{subequations}
\subsection{Third order contributions}
The contribution of $B^{(3)}\left(\rho_S,t_{1}\right)$ to the system master equation can be calculated by noting that only $\{123\}$ term of Eq.~(\ref{eq:b3c}) is nonzero. The brace $\{123\}$ has 
a total of 64 terms; half of those are equal to zero because they contain odd moments. Combination of eight triple products of the diamagnetic term $A^2(t_i)$ as well as eight terms formed by different permutations of $A^{2}\left(t_{1}\right)A\left(t_{2}\right)A\left(t_{3}\right)$  
cancel each other. Hence, the remaining 16 terms can be combined as four different integrals as $T_{1}^{(3)}=T_{31}+T_{33}$ and $T_{2}^{(3)}=T_{32}+T_{34}$, where 
\begin{subequations}
\begin{eqnarray}
T_{31}&=&\left(\frac{4\,\omega_{0}}{\hbar^{2}\,m^{2}}\right)\int_{0}^{\infty}dt_{12}\int_{0}^{\infty}dt_{23}\,\textrm{Im}\{\langle 12\rangle\}\langle 23\rangle\cos\left(\omega_{0}\,t_{13}\right)\label{eq:t31},\\
T_{32}&=&\left(\frac{4\,\omega_{0}}{\hbar^{2}\,m^{2}}\right)\int_{0}^{\infty}dt_{12}\int_{0}^{\infty}dt_{23}\,\textrm{Im}\{\langle 12\rangle\}\langle 23\rangle\sin\left(\omega_{0}\,t_{13}\right)\label{eq:t32},\\
T_{33}&=&\left(\frac{4\,\omega_{0}}{\hbar^{2}\,m^{2}}\right)\int_{0}^{\infty}dt_{12}\int_{0}^{\infty}dt_{23}\,\textrm{Im}\{\langle 13\rangle \langle 23\rangle\}\cos\left(\omega_{0}\,t_{12}\right)\label{eq:t33},\\
T_{34}&=&\left(\frac{4\,\omega_{0}}{\hbar^{2}\,m^{2}}\right)\int_{0}^{\infty}dt_{12}\int_{0}^{\infty}dt_{23}\,\textrm{Im}\{\langle 13\rangle \langle 23\rangle\}\sin\left(\omega_{0}\,t_{12}\right)\label{eq:t34}.
\end{eqnarray}
\end{subequations}
\noindent One should note that when the diamagnetic term is neglected, 
the third order contribution would be zero, so the nonzero third order 
contributions here are due to interactions that contain $A^{2}(t_2)$ and  $A^{2}(t_3)$. These integrals are discussed in the main body of the text.
\subsection{Fourth order contributions}
Evaluation of $B^{\left(4\right)}\left(\rho_{S},t_{1}\right)$, given in Eq.~(\ref{eq:b4d}), is more involved and tedious compared to that of the lower order terms. Luckily, most of the braces in Eq.~(\ref{eq:b4d}) contain a single element separated with columns and those are zero because of the Gaussian character of the bath. So, the nonzero fourth order terms can be written as:
\begin{eqnarray}
\label{B4eq}
B^{\left(4\right)}\left(\rho_{S},t_{1}\right)&=&\frac{1}{\left(i\,\hbar\right)^4}\int_{-\infty}^{t_{1}}dt_{2}\int_{-\infty}^{t_{2}}dt_{3}\int_{-\infty}^{t_{3}}dt_{4}\left\{\{1234\}-\{12:34\}-\{13:24\}\right.\nonumber\\
&&-\left.\{14:23\}\right\},
\end{eqnarray}
\noindent where the brace $\{1234\}$ has a total of 256 terms; half of those are zero because 
they involve the odd powers of the vector potential $A(t_i)$. Furthermore, 16 (48) terms which 
are in the form of product of four diamagnetic $A(t_i)^2$ (vector potential $A(t_1)^{2}\dots$) terms at 
times $t_1,\,t_2,\,t_3,\,t_4$ add up to zero. The remaining terms can be written as four and six 
time averages as: 
\begin{eqnarray}\label{full1234}
\{1234\}&=&\frac{1}{m^{4}}\left[\langle 4213\rangle\left[p\left(t_{1}\right),p\left(t_{3}\right)\rho_{S}\,p\left(t_{4}\right)p\left(t_{2}\right)\right]+\langle 1234\rangle\left[p\left(t_{1}\right),p\left(t_{2}\right)p\left(t_{3}\right)p\left(t_{4}\right)\rho_{S}\right]\right.\nonumber\\
&&-\left.\langle 2134\rangle\left[p\left(t_{1}\right),p\left(t_{3}\right)p\left(t_{4}\right)\rho_{S}\,p\left(t_{2}\right)\right]-\langle 4321\rangle\left[p\left(t_{1}\right),\rho_{S}\,p\left(t_{4}\right)p\left(t_{3}\right)p\left(t_{2}\right)\right]\right.\nonumber\\
&&-\left.\langle 3124\rangle\left[p\left(t_{1}\right),p\left(t_{2}\right)p\left(t_{4}\right)\rho_{S}\,p\left(t_{3}\right)\right]-\langle 4123\rangle\left[p\left(t_{1}\right),p\left(t_{2}\right)p\left(t_{3}\right)\rho_{S}\,p\left(t_{4}\right)\right]\right.\nonumber\\
&&+\left.\langle 4312\rangle\left[p\left(t_{1}\right),p\left(t_{2}\right)\rho_{S}\,p\left(t_{4}\right)p\left(t_{3}\right)\right]+\langle 3214\rangle\left[p\left(t_{1}\right),p\left(t_{4}\right)\rho_{S}\,p\left(t_{3}\right)p\left(t_{2}\right)\right]\right]\nonumber\\
&&+\frac{1}{4\,m^{4}}\left[\left(\langle 122334\rangle-\langle 221334\rangle+\langle 332214\rangle-\langle 331224\rangle\right)\left[p\left(t_{1}\right),p\left(t_{4}\right)\rho_{S}\,\right]\right.\nonumber\\
&&+\left.\left(\langle 422133\rangle-\langle 433221\rangle+\langle 433122\rangle-\langle 412233\rangle\right)\left[p\left(t_{1}\right),\rho_{S}\,p\left(t_{4}\right)\right]\right.\nonumber\\
&&+\left.\left(\langle 122344\rangle-\langle 221344\rangle+\langle 442213\rangle-\langle 441223\rangle\right)\left[p\left(t_{1}\right),p\left(t_{3}\right)\rho_{S}\,\right]\right.\nonumber\\
&&+\left.\left(\langle 322144\rangle-\langle 443221\rangle+\langle 443122\rangle-\langle 312244\rangle\right)\left[p\left(t_{1}\right),\rho_{S}\,p\left(t_{3}\right)\right]\right.\nonumber\\
&&+\left.\left(\langle 123344\rangle-\langle 331244\rangle+\langle 443312\rangle-\langle 441233\rangle\right)\left[p\left(t_{1}\right),p\left(t_{2}\right)\rho_{S}\,\right]\right.\nonumber\\
&&+\left.\left(\langle 332144\rangle-\langle 213344\rangle+\langle 442133\rangle-\langle 443321\rangle\right)\left[p\left(t_{1}\right),\rho_{S}\,p\left(t_{2}\right)\right]\right],
\end{eqnarray}
\noindent where four-time terms are due to $p(t).A(t)$ interaction, while six-time terms originate from $p(t_i)A(t_i)A^{2}(t_j)..$ type couplings that involve both diamagnetic and velocity interactions. The remaining braces in Eq.~(\ref{eq:b4d}) involve only velocity-coupling terms and can be expressed as:
\begin{eqnarray}\label{1234}
\{12:34\}&=&\frac{1}{m^{4}}\left[\langle 12\rangle\langle34\rangle\left[p\left(t_{1}\right),p\left(t_{2}\right)\left[p\left(t_{3}\right),p\left(t_{4}\right)\rho_{S}\right]\right]\right.\nonumber\\
&&+\left.\langle 12\rangle\langle 43\rangle\left[p\left(t_{1}\right),p\left(t_{2}\right)\left[\rho_{S}p\left(t_{4}\right),p\left(t_{3}\right)\right]\right]\right.\nonumber\\
&&-\left.\langle 21\rangle\langle 34\rangle\left[p\left(t_{1}\right),\left[p\left(t_{3}\right),p\left(t_{4}\right)\rho_{S}\right]p\left(t_{2}\right)\right]\right.\nonumber\\
&&-\left.\langle 21\rangle\langle 43\rangle\left[p\left(t_{1}\right),\left[\rho_{S}p\left(t_{4}\right),p\left(t_{3}\right)\right]p\left(t_{2}\right)\right]\right],
\end{eqnarray}
\begin{eqnarray}\label{1324}
\{13:24\}&=&\frac{1}{m^{4}}\left[\langle 13\rangle\langle 24\rangle\left[p\left(t_{1}\right),p\left(t_{3}\right)\left[p\left(t_{2}\right),p\left(t_{4}\right)\rho_{S}\right]\right]\right.\nonumber\\
&&+\left.\langle 13\rangle\langle 42\rangle\left[p\left(t_{1}\right),p\left(t_{3}\right)\left[\rho_{S}p\left(t_{4}\right),p\left(t_{2}\right)\right]\right]\right.\nonumber\\
&&-\left.\langle 31\rangle\langle 24\rangle\left[p\left(t_{1}\right),\left[p\left(t_{2}\right),p\left(t_{4}\right)\rho_{S}\right]p\left(t_{3}\right)\right]\right.\nonumber\\
&&-\left.\langle 31\rangle\langle 24\rangle\left[p\left(t_{1}\right),\left[\rho_{S}p\left(t_{4}\right),p\left(t_{2}\right)\right]p\left(t_{3}\right)\right]\right],
\end{eqnarray}
\begin{eqnarray}\label{1423}
\{14:23\}&=&\frac{1}{m^{4}}\left[\langle 14\rangle\langle 23\rangle\left[p\left(t_{1}\right),p\left(t_{4}\right)\left[p\left(t_{2}\right),p\left(t_{3}\right)\rho_{S}\right]\right]\right.\nonumber\\
&&+\left.\langle 14\rangle\langle 32\rangle\left[p\left(t_{1}\right),p\left(t_{4}\right)\left[\rho_{S}p\left(t_{3}\right),p\left(t_{2}\right)\right]\right]\right.\nonumber\\
&&-\left.\langle 41\rangle\langle 23\rangle\left[p\left(t_{1}\right),\left[p\left(t_{2}\right),p\left(t_{3}\right)\rho_{S}\right]p\left(t_{4}\right)\right]\right.\nonumber\\
&&-\left.\langle 41\rangle\langle 32\rangle\left[p\left(t_{1}\right),\left[\rho_{S}p\left(t_{3}\right),p\left(t_{2}\right)\right]p\left(t_{4}\right)\right]\right].
\end{eqnarray} 
Since Eq.~(\ref{full1234}) contains both four and six index terms, one needs to transform them into pair product form by using Isserlis' theorem. The formula for the four index term was given in Eq.~(\ref{even}) while the six index term can be written as: 
\begin{eqnarray}
\langle 123456\rangle&=&\langle 12\rangle \langle 34\rangle\langle 56\rangle+\langle 12\rangle \langle 35\rangle\langle 46\rangle +\langle 12\rangle \langle 36\rangle\langle 45\rangle\nonumber\\
&&+\langle 13\rangle \langle 24\rangle\langle 56\rangle+\langle 13\rangle \langle 25\rangle\langle 46\rangle+\langle 13\rangle \langle 26\rangle\langle 45\rangle\nonumber\\
&&+\langle 14\rangle \langle 23\rangle\langle 56\rangle+\langle 14\rangle \langle 25\rangle\langle 36\rangle+\langle 14\rangle \langle 26\rangle\langle 35\rangle\nonumber\\
&&+\langle 15\rangle \langle 23\rangle\langle 46\rangle+\langle 15\rangle \langle 24\rangle\langle 36\rangle+\langle 15\rangle \langle 26\rangle\langle 34\rangle\nonumber\\
&&+\langle 16\rangle \langle 23\rangle\langle 45\rangle+\langle 16\rangle \langle 24\rangle\langle 35\rangle+\langle 16\rangle \langle 25\rangle\langle 34\rangle.
\end{eqnarray}

In equations~(\ref{full1234})-(\ref{1423}), we can use the time dependent momentum operator~\cite{ford1996} 
\begin{equation}
p\left(t_2\right)=p\left(t_1\right)\,\cos{\left[\omega_0\left(t_1-t_2\right)\right]}+
m\,\omega_0\,x\left(t_1\right)\sin{\left[\omega_0\,\left(t_1-t_2\right)\right]}
\label{eq:ptime}
\end{equation}
\noindent to simplify the commutators involving the system momentum at different times and the system density matrix and obtain the contribution of $B^{(4)}\left(\rho_S,t_{1}\right)$ to the master equation 
$T^{(4)}_{1}=T_{41}+T_{61}+T_{63}$ and $T^{(4)}_{2}=T_{42}+T_{62}+T_{64}$
where 
\begin{eqnarray}
T_{41}&=&i\left(\frac{8\,\omega_{0}^{2}}{\hbar^2\, m^2}\right)\int_{0}^{\infty}dt_{12}\int_{0}^{\infty}dt_{23}\int_{0}^{\infty}dt_{34}\,\left\{\textrm{Im}
\left[\langle 14\rangle\right]\langle 23\rangle \cos{\left(\omega_{0}\,t_{13}\right)}
\sin{\left(\omega_{0}\,t_{24}\right)}\right.
\nonumber\\
&&+\left.\textrm{Im}\left[\langle 13\rangle\right]\langle 24\rangle\cos{\left(\omega_{0}\,t_{14}\right)}
\sin{\left(\omega_{0}\,t_{23}\right)}+
\textrm{Im}\left[\langle 14\rangle\langle 23\rangle\right]
\cos{\left(\omega_{0}\,t_{12}\right)}\sin{\left(\omega_{0}\, t_{34}\right)}\right\},
\nonumber\\
T_{42}&=&i\left(\frac{8\,\omega_{0}^{2}}{\hbar^2\, m^2}\right)\int_{0}^{\infty}dt_{12}\int_{0}^{\infty}dt_{23}\int_{0}^{\infty}dt_{34}\,\left\{\textrm{Im}
\left[\langle 14\rangle\right]\langle 23\rangle \sin{\left(\omega_{0}\,t_{13}\right)}
\sin{\left(\omega_{0}\,t_{24}\right)}\right. 
\nonumber\\
&&+\left.\textrm{Im}\left[\langle 13\rangle\right]\langle 24\rangle\sin{\left(\omega_{0}\,t_{14}\right)}
\sin{\left(\omega_{0}\,t_{23}\right)}+
\textrm{Im}\left[\langle 14\rangle\langle 23\rangle\right]\sin{\left(\omega_{0}\, t_{12}\right)}
\sin{\left(\omega_{0}\,t_{34}\right)}\right\},
\nonumber\\
T_{61}&=&\left(\frac{32\,\omega_{0}}{\hbar^3\, m^3}\right)\int_{0}^{\infty}dt_{12}\int_{0}^{\infty}dt_{23}\int_{0}^{\infty}dt_{34}\,
\left\{\textrm{Im}\left[\langle 12\rangle\right]
\textrm{Im}\left[\langle 23\rangle\right]\langle 34\rangle\cos{\left(\omega_{0}t_{14}\right)}\right\},
\nonumber\\
T_{62}&=&\left(\frac{32\omega_{0}}{\hbar^3 m^3}\right)\int_{0}^{\infty}dt_{12}\int_{0}^{\infty}dt_{23}\int_{0}^{\infty}dt_{34}\,
\left\{\textrm{Im}\left[\langle 12\rangle\right]
\textrm{Im}\left[\langle 23\rangle\right]\langle 34\rangle\sin{\left(\omega_{0}\,t_{14}\right)}\right\},
\nonumber\\
T_{63}&=&\left(\frac{32\,\omega_{0}}{\hbar^3\, m^3}\right)\int_{0}^{\infty}dt_{12}\int_{0}^{\infty}dt_{23}\int_{0}^{\infty}dt_{34}\,
\left\{\textrm{Im}\left[\langle 12\rangle\right]
\textrm{Im}\left[\langle 24\rangle\langle 34\rangle\right]\cos{\left(\omega_{0}\,t_{13}\right)}\right.
\nonumber\\
&&+\left.\left(\textrm{Im}\left[\langle 23\rangle\right]
\textrm{Im}\left[\langle 14\rangle\langle 34\rangle\right]
+\textrm{Im}\left[\langle 13\rangle\right]
\textrm{Im}\left[\langle 24\rangle\langle 34\rangle\right]\right)\cos{\left(\omega_{0}\,t_{12}\right)}
\right\},
\nonumber\\
T_{64}&=&\left(\frac{32\,\omega_{0}^2}{\hbar^3\, m^2}\right)\int_{0}^{\infty}dt_{12}\int_{0}^{\infty}dt_{23}\int_{0}^{\infty}dt_{34}\,
\left\{\textrm{Im}\left[\langle 12\rangle\right]
\textrm{Im}\left[\langle 24\rangle\langle 34\rangle\right]
\sin{\left(\omega_{0}\,t_{13}\right)}\right.
\nonumber\\
&&+\left.\left(\textrm{Im}
\left[\langle 23\rangle\right]
\textrm{Im}\left[\langle 14\rangle\langle 34\rangle\right]
+\textrm{Im}
\left[\langle 13\rangle\right]
\textrm{Im}\left[\langle 24\rangle\langle 34\rangle\right]\right)
\sin{\left(\omega_{0}\,t_{12}\right)}
\right\}.\label{eq:t46int}
\end{eqnarray}
Here $T_{6i}$, where $i=1,2,3,$ and $4$, contain the diamagnetic interaction terms while $T_{41}$ and $T_{42}$ are due to $p.A$ type interaction. Although tedious, all of $T_{4i}$ and $T_{6i}$ integrals can be carried out exactly by using the rational expansion of the hyperbolic cotangent function, as was the case for the second and the third order contributions. The results are:
\begin{subequations}
\begin{eqnarray}
T_{41}&=&-\gamma f_{\omega_0}^3\,\left[
\frac{1}{2}\left(\frac{\delta m}{m}-5 \frac{\gamma}{\Omega}\right)\coth{\left(
\beta\omega_0
\right)}+\frac{\gamma}{\omega_0}\,I
\right] \nonumber \\
&&-f^2_{\omega_0}\gamma\,\beta\left[
\frac{1}{\pi^2}\,\gamma\,\mathrm{Im}\left[\psi_1\left(\beta,\omega_0\right)\right]
-\frac{1}{2}\omega_0\frac{\delta m}{m}\csch^2\left(\beta\,\omega_{0}\right)
\right]\nonumber\\
&&-i\,\gamma^2\,f_{\omega_0}^3\left(\frac{1}{\omega_0}-\frac{\omega_0}{\Omega^2}\right) 
\label{eq:t41},\\
T_{42}&=&-\gamma^2\,f_{\omega_0}^3\left[\frac{1}{2}\left(\frac{\omega_0}{\Omega^2}-\frac{1}{\omega_0}\right)\coth{\left(\beta\,\omega_0\right)}+
\frac{1}{\beta\,\Omega^{2}}\left(1+\left(\frac{\omega_{0}}{\Omega}\right)^{2}\right)
-\frac{3}{\Omega}\,I\right]\nonumber\\
&&-f_{\omega_{0}}^{2}\,\gamma\,\beta\left[\gamma\,\left(
\frac{1}{2}\,\csch^2{\left(\beta\,\omega_{0}\right)}-\frac{2}{\pi^{2}}\,\psi_{1}\left[
\frac{\beta\,\Omega}{\pi}\right]
\right)+\frac{1}{\pi^{2}}\,\omega_{0}\,\frac{\delta m}{m}\,\mathrm{Im}\left[\psi_1\left(\beta,\omega_0\right)\right]\right]\nonumber\\
&&-i\,2\,f_{\omega_{0}}^{3}\frac{\gamma^{2}}{\Omega}\label{eq:t42},\\
T_{61}&=&-f_{\omega_{0}}^{3}\gamma\,\frac{\delta m}{m}\left[\left(\frac{\gamma}{\Omega}-\frac{\delta m}{m}\right)\,\coth{\left(\beta\,\omega_{0}\right)}+2\frac{\gamma}{\omega_0}\,I
\right]\nonumber\\
&&-i\,f_{\omega_{0}}^{3}\,\gamma\,\frac{\delta m}{m}\,\left[\frac{\omega_{0}}{\gamma}\,\left(\frac{\delta m}{m}\right)^{2}-3\,\frac{\gamma}{\omega_{0}}\right]\label{eq:t61},\\
T_{62}&=&-f_{\omega_{0}}^{3}\,\gamma\,\frac{\delta m}{m}\left[
\left(\frac{\gamma}{\Omega}-\frac{\delta m}{m}\right)\,I-2\frac{\gamma}{\omega_0}\coth{\left(
\beta\,\omega_0\right)}\right]\nonumber \\
&&-i\,f_{\omega_{0}}^{3}\,\gamma\,\frac{\delta m}{m}\left[3\,\frac{\delta m}{m}-\frac{\gamma}{\Omega}\right]\label{eq:t62},\\
T_{63}&=&-2f_{\omega_{0}}^{3}\,\gamma\,\frac{\delta m}{m}\,\left[
-\frac{\gamma}{\omega_0}\,I-
\frac{\delta m}{m}\coth{\left(\beta\omega_0\right)}
\right]\label{eq:t63},\\
T_{64}&=&-2f_{\omega_0}^3\gamma\frac{\delta m}{m}\left[\frac{\gamma}{\omega_0}
\coth{\left(\beta\omega_0\right)}-\frac{\delta m}{m}\,I
\right]\nonumber\\
&&-f_{\omega_0}\gamma\frac{\delta m}{m}\left[
\frac{f_{\omega_0}}{\beta\,\Omega}
\left(3\frac{\delta m}{m}+\frac{\gamma}{\Omega}
\right)+\frac{\delta m}{m}\,\beta\,\Omega\left(\frac{\beta\,\Omega}{\pi^3}\psi_2-\frac{4\,f_{\omega_0}}{\pi^2}\psi_1\right)
\right].
\end{eqnarray}
\end{subequations}


\begin{thebibliography}{44}%
\makeatletter
\providecommand \@ifxundefined [1]{%
 \@ifx{#1\undefined}
}%
\providecommand \@ifnum [1]{%
 \ifnum #1\expandafter \@firstoftwo
 \else \expandafter \@secondoftwo
 \fi
}%
\providecommand \@ifx [1]{%
 \ifx #1\expandafter \@firstoftwo
 \else \expandafter \@secondoftwo
 \fi
}%
\providecommand \natexlab [1]{#1}%
\providecommand \enquote  [1]{``#1''}%
\providecommand \bibnamefont  [1]{#1}%
\providecommand \bibfnamefont [1]{#1}%
\providecommand \citenamefont [1]{#1}%
\providecommand \href@noop [0]{\@secondoftwo}%
\providecommand \href [0]{\begingroup \@sanitize@url \@href}%
\providecommand \@href[1]{\@@startlink{#1}\@@href}%
\providecommand \@@href[1]{\endgroup#1\@@endlink}%
\providecommand \@sanitize@url [0]{\catcode `\\12\catcode `\$12\catcode
  `\&12\catcode `\#12\catcode `\^12\catcode `\_12\catcode `\%12\relax}%
\providecommand \@@startlink[1]{}%
\providecommand \@@endlink[0]{}%
\providecommand \url  [0]{\begingroup\@sanitize@url \@url }%
\providecommand \@url [1]{\endgroup\@href {#1}{\urlprefix }}%
\providecommand \urlprefix  [0]{URL }%
\providecommand \Eprint [0]{\href }%
\providecommand \doibase [0]{http://dx.doi.org/}%
\providecommand \selectlanguage [0]{\@gobble}%
\providecommand \bibinfo  [0]{\@secondoftwo}%
\providecommand \bibfield  [0]{\@secondoftwo}%
\providecommand \translation [1]{[#1]}%
\providecommand \BibitemOpen [0]{}%
\providecommand \bibitemStop [0]{}%
\providecommand \bibitemNoStop [0]{.\EOS\space}%
\providecommand \EOS [0]{\spacefactor3000\relax}%
\providecommand \BibitemShut  [1]{\csname bibitem#1\endcsname}%
\let\auto@bib@innerbib\@empty
%</preamble>
\bibitem [{\citenamefont {Lewenstein}\ \emph {et~al.}(2012)\citenamefont
  {Lewenstein}, \citenamefont {Sanpera},\ and\ \citenamefont
  {Ahunger}}]{lewenstein2012}%
  \BibitemOpen
  \bibfield  {author} {\bibinfo {author} {\bibfnamefont {M.}~\bibnamefont
  {Lewenstein}}, \bibinfo {author} {\bibfnamefont {A.}~\bibnamefont {Sanpera}},
  \ and\ \bibinfo {author} {\bibfnamefont {V.}~\bibnamefont {Ahunger}},\
  }\href@noop {} {\emph {\bibinfo {title} {Ultracold atoms in optical lattices:
  Simulating quantum many-body systems}}}\ (\bibinfo  {publisher} {Oxford
  University Press},\ \bibinfo {year} {2012})\BibitemShut {NoStop}%
\bibitem [{\citenamefont {Massignan}\ \emph {et~al.}(2015)\citenamefont
  {Massignan}, \citenamefont {Lampo}, \citenamefont {Wehr},\ and\ \citenamefont
  {Lewenstein}}]{massignan2015}%
  \BibitemOpen
  \bibfield  {author} {\bibinfo {author} {\bibfnamefont {P.}~\bibnamefont
  {Massignan}}, \bibinfo {author} {\bibfnamefont {A.}~\bibnamefont {Lampo}},
  \bibinfo {author} {\bibfnamefont {J.}~\bibnamefont {Wehr}}, \ and\ \bibinfo
  {author} {\bibfnamefont {M.}~\bibnamefont {Lewenstein}},\ }\href {\doibase
  10.1103/PhysRevA.91.033627} {\bibfield  {journal} {\bibinfo  {journal} {Phys.
  Rev. A}\ }\textbf {\bibinfo {volume} {91}},\ \bibinfo {pages} {033627}
  (\bibinfo {year} {2015})}\BibitemShut {NoStop}%
\bibitem [{\citenamefont {Breuer}\ and\ \citenamefont
  {Petruccione}(2002)}]{breuer2002theory}%
  \BibitemOpen
  \bibfield  {author} {\bibinfo {author} {\bibfnamefont {H.-P.}\ \bibnamefont
  {Breuer}}\ and\ \bibinfo {author} {\bibfnamefont {F.}~\bibnamefont
  {Petruccione}},\ }\href@noop {} {\emph {\bibinfo {title} {The theory of open
  quantum systems}}}\ (\bibinfo  {publisher} {Oxford University Press},\
  \bibinfo {year} {2002})\BibitemShut {NoStop}%
\bibitem [{\citenamefont {Schlosshauer}(2007)}]{schlosshauer2007decoherence}%
  \BibitemOpen
  \bibfield  {author} {\bibinfo {author} {\bibfnamefont {M.~A.}\ \bibnamefont
  {Schlosshauer}},\ }\href@noop {} {\emph {\bibinfo {title} {Decoherence: and
  the quantum-to-classical transition}}}\ (\bibinfo  {publisher} {Springer
  Science \& Business Media},\ \bibinfo {year} {2007})\BibitemShut {NoStop}%
\bibitem [{\citenamefont {Weiss}(2008)}]{weiss1999quantum}%
  \BibitemOpen
  \bibfield  {author} {\bibinfo {author} {\bibfnamefont {U.}~\bibnamefont
  {Weiss}},\ }\href@noop {} {\emph {\bibinfo {title} {Quantum dissipative
  systems}}}\ (\bibinfo  {publisher} {World Scientific},\ \bibinfo {year}
  {2008})\BibitemShut {NoStop}%
\bibitem [{\citenamefont {Hu}\ \emph {et~al.}(1992)\citenamefont {Hu},
  \citenamefont {Paz},\ and\ \citenamefont {Zhang}}]{hu1992}%
  \BibitemOpen
  \bibfield  {author} {\bibinfo {author} {\bibfnamefont {B.~L.}\ \bibnamefont
  {Hu}}, \bibinfo {author} {\bibfnamefont {J.~P.}\ \bibnamefont {Paz}}, \ and\
  \bibinfo {author} {\bibfnamefont {Y.}~\bibnamefont {Zhang}},\ }\href@noop {}
  {\bibfield  {journal} {\bibinfo  {journal} {Phys. Rev. D}\ }\textbf {\bibinfo
  {volume} {45}},\ \bibinfo {pages} {2843} (\bibinfo {year}
  {1992})}\BibitemShut {NoStop}%
\bibitem [{\citenamefont {Tanimura}(2006)}]{tanimura2006}%
  \BibitemOpen
  \bibfield  {author} {\bibinfo {author} {\bibfnamefont {Y.}~\bibnamefont
  {Tanimura}},\ }\href@noop {} {\bibfield  {journal} {\bibinfo  {journal} {J.
  Phys. Soc. Jpn.}\ }\textbf {\bibinfo {volume} {75}},\ \bibinfo {pages}
  {082001} (\bibinfo {year} {2006})}\BibitemShut {NoStop}%
\bibitem [{\citenamefont {Liu}\ \emph {et~al.}(2014)\citenamefont {Liu},
  \citenamefont {Zhu}, \citenamefont {Bai},\ and\ \citenamefont
  {Shi}}]{liu2014}%
  \BibitemOpen
  \bibfield  {author} {\bibinfo {author} {\bibfnamefont {H.}~\bibnamefont
  {Liu}}, \bibinfo {author} {\bibfnamefont {L.}~\bibnamefont {Zhu}}, \bibinfo
  {author} {\bibfnamefont {S.}~\bibnamefont {Bai}}, \ and\ \bibinfo {author}
  {\bibfnamefont {Q.}~\bibnamefont {Shi}},\ }\href {\doibase
  http://dx.doi.org/10.1063/1.4870035} {\bibfield  {journal} {\bibinfo
  {journal} {J. Chem. Phys.}\ }\textbf {\bibinfo {volume} {140}},\ \bibinfo
  {eid} {134106} (\bibinfo {year} {2014})}\BibitemShut {NoStop}%
\bibitem [{\citenamefont {Fleming}\ and\ \citenamefont
  {Cummings}(2011)}]{fleming2011}%
  \BibitemOpen
  \bibfield  {author} {\bibinfo {author} {\bibfnamefont {C.~H.}\ \bibnamefont
  {Fleming}}\ and\ \bibinfo {author} {\bibfnamefont {N.~I.}\ \bibnamefont
  {Cummings}},\ }\href {\doibase 10.1103/PhysRevE.83.031117} {\bibfield
  {journal} {\bibinfo  {journal} {Phys. Rev. E}\ }\textbf {\bibinfo {volume}
  {83}},\ \bibinfo {pages} {031117} (\bibinfo {year} {2011})}\BibitemShut
  {NoStop}%
\bibitem [{\citenamefont {Budimir}\ and\ \citenamefont
  {Skinner}(1987)}]{Skinner87}%
  \BibitemOpen
  \bibfield  {author} {\bibinfo {author} {\bibfnamefont {J.}~\bibnamefont
  {Budimir}}\ and\ \bibinfo {author} {\bibfnamefont {J.~L.}\ \bibnamefont
  {Skinner}},\ }\href@noop {} {\bibfield  {journal} {\bibinfo  {journal} {J.
  Stat. Phys.}\ }\textbf {\bibinfo {volume} {49}},\ \bibinfo {pages} {5}
  (\bibinfo {year} {1987})}\BibitemShut {NoStop}%
\bibitem [{\citenamefont {Laird}\ \emph {et~al.}(1991)\citenamefont {Laird},
  \citenamefont {Budimir},\ and\ \citenamefont {Skinner}}]{laird1991}%
  \BibitemOpen
  \bibfield  {author} {\bibinfo {author} {\bibfnamefont {B.~B.}\ \bibnamefont
  {Laird}}, \bibinfo {author} {\bibfnamefont {J.}~\bibnamefont {Budimir}}, \
  and\ \bibinfo {author} {\bibfnamefont {J.~L.}\ \bibnamefont {Skinner}},\
  }\href {\doibase http://dx.doi.org/10.1063/1.460626} {\bibfield  {journal}
  {\bibinfo  {journal} {J. Chem. Phys.}\ }\textbf {\bibinfo {volume} {94}},\
  \bibinfo {pages} {4391} (\bibinfo {year} {1991})}\BibitemShut {NoStop}%
\bibitem [{\citenamefont {Reichman}\ \emph {et~al.}(1997)\citenamefont
  {Reichman}, \citenamefont {Brown},\ and\ \citenamefont
  {Neu}}]{PhysRevE.55.2328}%
  \BibitemOpen
  \bibfield  {author} {\bibinfo {author} {\bibfnamefont {D.~R.}\ \bibnamefont
  {Reichman}}, \bibinfo {author} {\bibfnamefont {F.~L.~H.}\ \bibnamefont
  {Brown}}, \ and\ \bibinfo {author} {\bibfnamefont {P.}~\bibnamefont {Neu}},\
  }\href {\doibase 10.1103/PhysRevE.55.2328} {\bibfield  {journal} {\bibinfo
  {journal} {Phys. Rev. E}\ }\textbf {\bibinfo {volume} {55}},\ \bibinfo
  {pages} {2328} (\bibinfo {year} {1997})}\BibitemShut {NoStop}%
\bibitem [{\citenamefont {Thingna}\ \emph {et~al.}(2014)\citenamefont
  {Thingna}, \citenamefont {Zhou},\ and\ \citenamefont {Wang}}]{1408}%
  \BibitemOpen
  \bibfield  {author} {\bibinfo {author} {\bibfnamefont {J.}~\bibnamefont
  {Thingna}}, \bibinfo {author} {\bibfnamefont {H.}~\bibnamefont {Zhou}}, \
  and\ \bibinfo {author} {\bibfnamefont {J.~S.}\ \bibnamefont {Wang}},\
  }\href@noop {} {\bibfield  {journal} {\bibinfo  {journal} {J. Chem. Phys.}\
  }\textbf {\bibinfo {volume} {141}},\ \bibinfo {pages} {194101} (\bibinfo
  {year} {2014})}\BibitemShut {NoStop}%
\bibitem [{\citenamefont {Ford}\ \emph {et~al.}(1996)\citenamefont {Ford},
  \citenamefont {Lewis},\ and\ \citenamefont {O'Connell}}]{ford1996}%
  \BibitemOpen
  \bibfield  {author} {\bibinfo {author} {\bibfnamefont {G.~W.}\ \bibnamefont
  {Ford}}, \bibinfo {author} {\bibfnamefont {J.~T.}\ \bibnamefont {Lewis}}, \
  and\ \bibinfo {author} {\bibfnamefont {R.~F.}\ \bibnamefont {O'Connell}},\
  }\href@noop {} {\bibfield  {journal} {\bibinfo  {journal} {Ann. Phys.}\
  }\textbf {\bibinfo {volume} {252}},\ \bibinfo {pages} {362} (\bibinfo {year}
  {1996})}\BibitemShut {NoStop}%
\bibitem [{\citenamefont {Reichman}\ and\ \citenamefont
  {Silbey}(1996)}]{reichman1996}%
  \BibitemOpen
  \bibfield  {author} {\bibinfo {author} {\bibfnamefont {D.~R.}\ \bibnamefont
  {Reichman}}\ and\ \bibinfo {author} {\bibfnamefont {R.~J.}\ \bibnamefont
  {Silbey}},\ }\href@noop {} {\bibfield  {journal} {\bibinfo  {journal} {J.
  Chem. Phys.}\ }\textbf {\bibinfo {volume} {104}},\ \bibinfo {pages} {1506}
  (\bibinfo {year} {1996})}\BibitemShut {NoStop}%
\bibitem [{\citenamefont {Breuer}\ \emph
  {et~al.}(1999{\natexlab{a}})\citenamefont {Breuer}, \citenamefont {Faller},
  \citenamefont {Kappler},\ and\ \citenamefont {Petruccione}}]{Breuer1999}%
  \BibitemOpen
  \bibfield  {author} {\bibinfo {author} {\bibfnamefont {H.~P.}\ \bibnamefont
  {Breuer}}, \bibinfo {author} {\bibfnamefont {D.}~\bibnamefont {Faller}},
  \bibinfo {author} {\bibfnamefont {B.}~\bibnamefont {Kappler}}, \ and\
  \bibinfo {author} {\bibfnamefont {F.}~\bibnamefont {Petruccione}},\ }\href
  {\doibase 10.1103/PhysRevA.60.3188} {\bibfield  {journal} {\bibinfo
  {journal} {Phys. Rev. A}\ }\textbf {\bibinfo {volume} {60}},\ \bibinfo
  {pages} {3188} (\bibinfo {year} {1999}{\natexlab{a}})}\BibitemShut {NoStop}%
\bibitem [{\citenamefont {Breuer}\ \emph
  {et~al.}(1999{\natexlab{b}})\citenamefont {Breuer}, \citenamefont {Kappler},\
  and\ \citenamefont {Petruccione}}]{Breuer19992}%
  \BibitemOpen
  \bibfield  {author} {\bibinfo {author} {\bibfnamefont {H.~P.}\ \bibnamefont
  {Breuer}}, \bibinfo {author} {\bibfnamefont {B.}~\bibnamefont {Kappler}}, \
  and\ \bibinfo {author} {\bibfnamefont {F.}~\bibnamefont {Petruccione}},\
  }\href {\doibase 10.1103/PhysRevA.59.1633} {\bibfield  {journal} {\bibinfo
  {journal} {Phys. Rev. A}\ }\textbf {\bibinfo {volume} {59}},\ \bibinfo
  {pages} {1633} (\bibinfo {year} {1999}{\natexlab{b}})}\BibitemShut {NoStop}%
\bibitem [{\citenamefont {Breuer}\ \emph {et~al.}(2001)\citenamefont {Breuer},
  \citenamefont {Kappler},\ and\ \citenamefont {Petruccione}}]{breuer200136}%
  \BibitemOpen
  \bibfield  {author} {\bibinfo {author} {\bibfnamefont {H.~P.}\ \bibnamefont
  {Breuer}}, \bibinfo {author} {\bibfnamefont {B.}~\bibnamefont {Kappler}}, \
  and\ \bibinfo {author} {\bibfnamefont {F.}~\bibnamefont {Petruccione}},\
  }\href {\doibase http://dx.doi.org/10.1006/aphy.2001.6152} {\bibfield
  {journal} {\bibinfo  {journal} {Ann. Phys.}\ }\textbf {\bibinfo {volume}
  {291}},\ \bibinfo {pages} {36 } (\bibinfo {year} {2001})}\BibitemShut
  {NoStop}%
\bibitem [{\citenamefont {Jang}\ \emph {et~al.}(2002)\citenamefont {Jang},
  \citenamefont {Cao},\ and\ \citenamefont {Silbey}}]{jang2002}%
  \BibitemOpen
  \bibfield  {author} {\bibinfo {author} {\bibfnamefont {S.}~\bibnamefont
  {Jang}}, \bibinfo {author} {\bibfnamefont {J.}~\bibnamefont {Cao}}, \ and\
  \bibinfo {author} {\bibfnamefont {R.~J.}\ \bibnamefont {Silbey}},\ }\href
  {\doibase http://dx.doi.org/10.1063/1.1445105} {\bibfield  {journal}
  {\bibinfo  {journal} {J. Chem. Phys.}\ }\textbf {\bibinfo {volume} {116}},\
  \bibinfo {pages} {2705} (\bibinfo {year} {2002})}\BibitemShut {NoStop}%
\bibitem [{\citenamefont {Doll}\ \emph {et~al.}(2008)\citenamefont {Doll},
  \citenamefont {Zueco}, \citenamefont {Wubs}, \citenamefont {Kohler},\ and\
  \citenamefont {Hanggi}}]{doll2008243}%
  \BibitemOpen
  \bibfield  {author} {\bibinfo {author} {\bibfnamefont {R.}~\bibnamefont
  {Doll}}, \bibinfo {author} {\bibfnamefont {D.}~\bibnamefont {Zueco}},
  \bibinfo {author} {\bibfnamefont {M.}~\bibnamefont {Wubs}}, \bibinfo {author}
  {\bibfnamefont {S.}~\bibnamefont {Kohler}}, \ and\ \bibinfo {author}
  {\bibfnamefont {P.}~\bibnamefont {Hanggi}},\ }\href {\doibase
  http://dx.doi.org/10.1016/j.chemphys.2007.09.003} {\bibfield  {journal}
  {\bibinfo  {journal} {Chem. Phys.}\ }\textbf {\bibinfo {volume} {347}},\
  \bibinfo {pages} {243 } (\bibinfo {year} {2008})}\BibitemShut {NoStop}%
\bibitem [{\citenamefont {Ishizaki}\ and\ \citenamefont
  {Tanimura}(2008)}]{tanimura2008}%
  \BibitemOpen
  \bibfield  {author} {\bibinfo {author} {\bibfnamefont {A.}~\bibnamefont
  {Ishizaki}}\ and\ \bibinfo {author} {\bibfnamefont {Y.}~\bibnamefont
  {Tanimura}},\ }\href {\doibase
  http://dx.doi.org/10.1016/j.chemphys.2007.10.037} {\bibfield  {journal}
  {\bibinfo  {journal} {Chem. Phys.}\ }\textbf {\bibinfo {volume} {347}},\
  \bibinfo {pages} {185 } (\bibinfo {year} {2008})}\BibitemShut {NoStop}%
\bibitem [{\citenamefont {Singh}\ and\ \citenamefont
  {Brumer}(2012)}]{singh2012}%
  \BibitemOpen
  \bibfield  {author} {\bibinfo {author} {\bibfnamefont {N.}~\bibnamefont
  {Singh}}\ and\ \bibinfo {author} {\bibfnamefont {P.}~\bibnamefont {Brumer}},\
  }\href {\doibase 10.1080/00268976.2012.683457} {\bibfield  {journal}
  {\bibinfo  {journal} {Mol. Phys.}\ }\textbf {\bibinfo {volume} {110}},\
  \bibinfo {pages} {1815} (\bibinfo {year} {2012})}\BibitemShut {NoStop}%
\bibitem [{\citenamefont {Mavros}\ and\ \citenamefont
  {Voorhis}(2014)}]{Mavros2014}%
  \BibitemOpen
  \bibfield  {author} {\bibinfo {author} {\bibfnamefont {M.~G.}\ \bibnamefont
  {Mavros}}\ and\ \bibinfo {author} {\bibfnamefont {T.~V.}\ \bibnamefont
  {Voorhis}},\ }\href@noop {} {\bibfield  {journal} {\bibinfo  {journal} {J.
  Chem. Phys.}\ }\textbf {\bibinfo {volume} {141}},\ \bibinfo {pages} {054112}
  (\bibinfo {year} {2014})}\BibitemShut {NoStop}%
\bibitem [{\citenamefont {Ford}\ \emph
  {et~al.}(1988{\natexlab{a}})\citenamefont {Ford}, \citenamefont {Lewis},\
  and\ \citenamefont {O'Connell}}]{ford88}%
  \BibitemOpen
  \bibfield  {author} {\bibinfo {author} {\bibfnamefont {G.~W.}\ \bibnamefont
  {Ford}}, \bibinfo {author} {\bibfnamefont {J.~T.}\ \bibnamefont {Lewis}}, \
  and\ \bibinfo {author} {\bibfnamefont {R.~F.}\ \bibnamefont {O'Connell}},\
  }\href {\doibase 10.1103/PhysRevA.37.4419} {\bibfield  {journal} {\bibinfo
  {journal} {Phys. Rev. A}\ }\textbf {\bibinfo {volume} {37}},\ \bibinfo
  {pages} {4419} (\bibinfo {year} {1988}{\natexlab{a}})}\BibitemShut {NoStop}%
\bibitem [{\citenamefont {Marathe}(1989)}]{marathe1989}%
  \BibitemOpen
  \bibfield  {author} {\bibinfo {author} {\bibfnamefont {Y.}~\bibnamefont
  {Marathe}},\ }\href@noop {} {\bibfield  {journal} {\bibinfo  {journal} {Phys.
  Rev. A}\ }\textbf {\bibinfo {volume} {39}},\ \bibinfo {pages} {5927}
  (\bibinfo {year} {1989})}\BibitemShut {NoStop}%
\bibitem [{\citenamefont {{Li, X. L. and Ford, G. W. and
  O'Connell}}(1996)}]{li1996}%
  \BibitemOpen
  \bibfield  {author} {\bibinfo {author} {\bibfnamefont {R.~F.}\ \bibnamefont
  {{Li, X. L. and Ford, G. W. and O'Connell}}},\ }\href@noop {} {\bibfield
  {journal} {\bibinfo  {journal} {Phys. Rev. E}\ }\textbf {\bibinfo {volume}
  {53}},\ \bibinfo {pages} {3359} (\bibinfo {year} {1996})}\BibitemShut
  {NoStop}%
\bibitem [{\citenamefont {Dattagupta}\ and\ \citenamefont
  {Singh}(1997)}]{dattagupta1997}%
  \BibitemOpen
  \bibfield  {author} {\bibinfo {author} {\bibfnamefont {S.}~\bibnamefont
  {Dattagupta}}\ and\ \bibinfo {author} {\bibfnamefont {J.}~\bibnamefont
  {Singh}},\ }\href {\doibase 10.1103/PhysRevLett.77.1413} {\bibfield
  {journal} {\bibinfo  {journal} {Phys. Rev. Lett.}\ }\textbf {\bibinfo
  {volume} {79}},\ \bibinfo {pages} {961} (\bibinfo {year} {1997})}\BibitemShut
  {NoStop}%
\bibitem [{\citenamefont {Bao}\ and\ \citenamefont {Bai}(2005)}]{bao2005}%
  \BibitemOpen
  \bibfield  {author} {\bibinfo {author} {\bibfnamefont {J.-D.}\ \bibnamefont
  {Bao}}\ and\ \bibinfo {author} {\bibfnamefont {Z.-W.}\ \bibnamefont {Bai}},\
  }\href {\doibase 10.1088/0256-307X/22/8/006} {\bibfield  {journal} {\bibinfo
  {journal} {Chin. Phys. Lett.}\ }\textbf {\bibinfo {volume} {22}},\ \bibinfo
  {pages} {1845} (\bibinfo {year} {2005})}\BibitemShut {NoStop}%
\bibitem [{\citenamefont {Bai}\ \emph {et~al.}(2005{\natexlab{a}})\citenamefont
  {Bai}, \citenamefont {Bao},\ and\ \citenamefont {Song}}]{bai2005}%
  \BibitemOpen
  \bibfield  {author} {\bibinfo {author} {\bibfnamefont {Z.~W.}\ \bibnamefont
  {Bai}}, \bibinfo {author} {\bibfnamefont {J.~D.}\ \bibnamefont {Bao}}, \ and\
  \bibinfo {author} {\bibfnamefont {Y.~L.}\ \bibnamefont {Song}},\ }\href
  {\doibase 10.1103/PhysRevE.72.061105} {\bibfield  {journal} {\bibinfo
  {journal} {Phys. Rev. E}\ }\textbf {\bibinfo {volume} {72}},\ \bibinfo
  {pages} {061105} (\bibinfo {year} {2005}{\natexlab{a}})}\BibitemShut
  {NoStop}%
\bibitem [{\citenamefont {Kalandarov}\ \emph {et~al.}(2007)\citenamefont
  {Kalandarov}, \citenamefont {Kanokov}, \citenamefont {Adamian},\ and\
  \citenamefont {Antonenko}}]{kalandarov2007}%
  \BibitemOpen
  \bibfield  {author} {\bibinfo {author} {\bibfnamefont {S.~A.}\ \bibnamefont
  {Kalandarov}}, \bibinfo {author} {\bibfnamefont {Z.}~\bibnamefont {Kanokov}},
  \bibinfo {author} {\bibfnamefont {G.~G.}\ \bibnamefont {Adamian}}, \ and\
  \bibinfo {author} {\bibfnamefont {N.~V.}\ \bibnamefont {Antonenko}},\ }\href
  {\doibase 10.1103/PhysRevE.75.031115} {\bibfield  {journal} {\bibinfo
  {journal} {Phys. Rev. E}\ }\textbf {\bibinfo {volume} {75}},\ \bibinfo
  {pages} {031115} (\bibinfo {year} {2007})}\BibitemShut {NoStop}%
\bibitem [{\citenamefont {Pach\'{o}n}\ and\ \citenamefont
  {Brumer}(2013)}]{pachon2013}%
  \BibitemOpen
  \bibfield  {author} {\bibinfo {author} {\bibfnamefont {L.~A.}\ \bibnamefont
  {Pach\'{o}n}}\ and\ \bibinfo {author} {\bibfnamefont {P.}~\bibnamefont
  {Brumer}},\ }\href {\doibase 10.1103/PhysRevA.87.022106} {\bibfield
  {journal} {\bibinfo  {journal} {Phys. Rev. A}\ }\textbf {\bibinfo {volume}
  {87}},\ \bibinfo {pages} {022106} (\bibinfo {year} {2013})}\BibitemShut
  {NoStop}%
\bibitem [{\citenamefont {Kumar}(2014)}]{kumar2014}%
  \BibitemOpen
  \bibfield  {author} {\bibinfo {author} {\bibfnamefont {J.}~\bibnamefont
  {Kumar}},\ }\href {\doibase 10.1016/j.physa.2013.08.046} {\bibfield
  {journal} {\bibinfo  {journal} {Physica A}\ }\textbf {\bibinfo {volume}
  {393}},\ \bibinfo {pages} {182} (\bibinfo {year} {2014})}\BibitemShut
  {NoStop}%
\bibitem [{\citenamefont {Bai}\ \emph {et~al.}(2005{\natexlab{b}})\citenamefont
  {Bai}, \citenamefont {Bao},\ and\ \citenamefont {Song}}]{Bai05}%
  \BibitemOpen
  \bibfield  {author} {\bibinfo {author} {\bibfnamefont {Z.-W.}\ \bibnamefont
  {Bai}}, \bibinfo {author} {\bibfnamefont {J.-D.}\ \bibnamefont {Bao}}, \ and\
  \bibinfo {author} {\bibfnamefont {Y.-L.}\ \bibnamefont {Song}},\ }\href
  {\doibase 10.1103/PhysRevE.72.061105} {\bibfield  {journal} {\bibinfo
  {journal} {Phys. Rev. E}\ }\textbf {\bibinfo {volume} {72}},\ \bibinfo
  {pages} {061105} (\bibinfo {year} {2005}{\natexlab{b}})}\BibitemShut
  {NoStop}%
\bibitem [{\citenamefont {Ford}\ \emph
  {et~al.}(1988{\natexlab{b}})\citenamefont {Ford}, \citenamefont {Lewis},\
  and\ \citenamefont {O'Connell}}]{ford1988}%
  \BibitemOpen
  \bibfield  {author} {\bibinfo {author} {\bibfnamefont {G.~W.}\ \bibnamefont
  {Ford}}, \bibinfo {author} {\bibfnamefont {J.~T.}\ \bibnamefont {Lewis}}, \
  and\ \bibinfo {author} {\bibfnamefont {R.~F.}\ \bibnamefont {O'Connell}},\
  }\href {\doibase 10.1103/PhysRevA.37.3609} {\bibfield  {journal} {\bibinfo
  {journal} {Phys. Rev. A}\ }\textbf {\bibinfo {volume} {37}},\ \bibinfo
  {pages} {3609} (\bibinfo {year} {1988}{\natexlab{b}})}\BibitemShut {NoStop}%
\bibitem [{\citenamefont {Babiker}\ \emph {et~al.}(1974)\citenamefont
  {Babiker}, \citenamefont {Power},\ and\ \citenamefont
  {Thirunamachandran}}]{Babiker1974}%
  \BibitemOpen
  \bibfield  {author} {\bibinfo {author} {\bibfnamefont {M.}~\bibnamefont
  {Babiker}}, \bibinfo {author} {\bibfnamefont {E.~A.}\ \bibnamefont {Power}},
  \ and\ \bibinfo {author} {\bibfnamefont {T.}~\bibnamefont
  {Thirunamachandran}},\ }\href@noop {} {\bibfield  {journal} {\bibinfo
  {journal} {Proc. R. Soc. A}\ }\textbf {\bibinfo {volume} {338}},\ \bibinfo
  {pages} {235} (\bibinfo {year} {1974})}\BibitemShut {NoStop}%
\bibitem [{\citenamefont {Castrigiano}\ and\ \citenamefont
  {Kokiantonis}(1987)}]{PhysRevA.35.4122}%
  \BibitemOpen
  \bibfield  {author} {\bibinfo {author} {\bibfnamefont {D.~P.~L.}\
  \bibnamefont {Castrigiano}}\ and\ \bibinfo {author} {\bibfnamefont
  {N.}~\bibnamefont {Kokiantonis}},\ }\href {\doibase 10.1103/PhysRevA.35.4122}
  {\bibfield  {journal} {\bibinfo  {journal} {Phys. Rev. A}\ }\textbf {\bibinfo
  {volume} {35}},\ \bibinfo {pages} {4122} (\bibinfo {year}
  {1987})}\BibitemShut {NoStop}%
\bibitem [{\citenamefont {Barone}\ and\ \citenamefont
  {Caldeira}(1991)}]{PhysRevA.43.57}%
  \BibitemOpen
  \bibfield  {author} {\bibinfo {author} {\bibfnamefont {P.~M. V.~B.}\
  \bibnamefont {Barone}}\ and\ \bibinfo {author} {\bibfnamefont {A.~O.}\
  \bibnamefont {Caldeira}},\ }\href {\doibase 10.1103/PhysRevA.43.57}
  {\bibfield  {journal} {\bibinfo  {journal} {Phys. Rev. A}\ }\textbf {\bibinfo
  {volume} {43}},\ \bibinfo {pages} {57} (\bibinfo {year} {1991})}\BibitemShut
  {NoStop}%
\bibitem [{\citenamefont {Castrigiano}\ and\ \citenamefont
  {Kokiantonis}(1988)}]{PhysRevA.38.527}%
  \BibitemOpen
  \bibfield  {author} {\bibinfo {author} {\bibfnamefont {D.~P.~L.}\
  \bibnamefont {Castrigiano}}\ and\ \bibinfo {author} {\bibfnamefont
  {N.}~\bibnamefont {Kokiantonis}},\ }\href {\doibase 10.1103/PhysRevA.38.527}
  {\bibfield  {journal} {\bibinfo  {journal} {Phys. Rev. A}\ }\textbf {\bibinfo
  {volume} {38}},\ \bibinfo {pages} {527} (\bibinfo {year} {1988})}\BibitemShut
  {NoStop}%
\bibitem [{\citenamefont {O'Connell}(2003)}]{OConnell2003}%
  \BibitemOpen
  \bibfield  {author} {\bibinfo {author} {\bibfnamefont {R.~F.}\ \bibnamefont
  {O'Connell}},\ }\href@noop {} {\bibfield  {journal} {\bibinfo  {journal}
  {Phys. Lett. A}\ }\textbf {\bibinfo {volume} {313}},\ \bibinfo {pages} {491}
  (\bibinfo {year} {2003})}\BibitemShut {NoStop}%
\bibitem [{\citenamefont {Chaturvedi}\ and\ \citenamefont
  {Shibata}(1979)}]{ZPhysB}%
  \BibitemOpen
  \bibfield  {author} {\bibinfo {author} {\bibfnamefont {S.}~\bibnamefont
  {Chaturvedi}}\ and\ \bibinfo {author} {\bibfnamefont {J.}~\bibnamefont
  {Shibata}},\ }\href@noop {} {\bibfield  {journal} {\bibinfo  {journal} {Z.
  Phys. B}\ }\textbf {\bibinfo {volume} {35}},\ \bibinfo {pages} {297}
  (\bibinfo {year} {1979})}\BibitemShut {NoStop}%
\bibitem [{\citenamefont {Yonn}\ \emph {et~al.}(1975)\citenamefont {Yonn},
  \citenamefont {Deutch},\ and\ \citenamefont {Freed}}]{Yoon75}%
  \BibitemOpen
  \bibfield  {author} {\bibinfo {author} {\bibfnamefont {B.}~\bibnamefont
  {Yonn}}, \bibinfo {author} {\bibfnamefont {J.~M.}\ \bibnamefont {Deutch}}, \
  and\ \bibinfo {author} {\bibfnamefont {J.~H.}\ \bibnamefont {Freed}},\
  }\href@noop {} {\bibfield  {journal} {\bibinfo  {journal} {J. Chem. Phys.}\
  }\textbf {\bibinfo {volume} {62}},\ \bibinfo {pages} {4687} (\bibinfo {year}
  {1975})}\BibitemShut {NoStop}%
\bibitem [{\citenamefont {Feynman}\ and\ \citenamefont
  {Vernon}(1963)}]{Feynman}%
  \BibitemOpen
  \bibfield  {author} {\bibinfo {author} {\bibfnamefont {R.~P.}\ \bibnamefont
  {Feynman}}\ and\ \bibinfo {author} {\bibfnamefont {F.~L.}\ \bibnamefont
  {Vernon}},\ }\href@noop {} {\bibfield  {journal} {\bibinfo  {journal} {Ann.
  Phys.(N.Y.)}\ }\textbf {\bibinfo {volume} {24}},\ \bibinfo {pages} {118}
  (\bibinfo {year} {1963})}\BibitemShut {NoStop}%
\bibitem [{\citenamefont {Fleming}\ \emph {et~al.}(2011)\citenamefont
  {Fleming}, \citenamefont {Roura},\ and\ \citenamefont
  {Hu}}]{fleming2011exact}%
  \BibitemOpen
  \bibfield  {author} {\bibinfo {author} {\bibfnamefont {C.}~\bibnamefont
  {Fleming}}, \bibinfo {author} {\bibfnamefont {A.}~\bibnamefont {Roura}}, \
  and\ \bibinfo {author} {\bibfnamefont {B.}~\bibnamefont {Hu}},\ }\href@noop
  {} {\bibfield  {journal} {\bibinfo  {journal} {Ann. Phys.}\ }\textbf
  {\bibinfo {volume} {326}},\ \bibinfo {pages} {1207} (\bibinfo {year}
  {2011})}\BibitemShut {NoStop}%
\bibitem [{\citenamefont {Paz}\ and\ \citenamefont
  {Roncaglia}(2009)}]{paz2009}%
  \BibitemOpen
  \bibfield  {author} {\bibinfo {author} {\bibfnamefont {J.~P.}\ \bibnamefont
  {Paz}}\ and\ \bibinfo {author} {\bibfnamefont {A.~J.}\ \bibnamefont
  {Roncaglia}},\ }\href {\doibase 10.1103/PhysRevA.79.032102} {\bibfield
  {journal} {\bibinfo  {journal} {Phys. Rev. A}\ }\textbf {\bibinfo {volume}
  {79}},\ \bibinfo {pages} {032102} (\bibinfo {year} {2009})}\BibitemShut
  {NoStop}%
\end{thebibliography}
\end{document}